%% file: ALRD_13CO_C18O_arxiv.tex
\documentclass[usenatbib]{mn2e}

\usepackage{psfig}
\usepackage{graphicx}

\def\apj{ApJ} \def\apjl{ApJL} \def\mnras{MNRAS} 
 \def\aaps{AAPS} \def\araa{ARAA} \def\aap{A\&A}
\def\aj{AJ} \def\apjs{APJS}  \def\nat{Nature}
\def\caa{ChA\&A}
\def\gs{\mathrel{\raise0.35ex\hbox{$\scriptstyle >$}\kern-0.6em\lower0.40ex\hbox{{$\scriptstyle \sim$}}}}
\def\ls{\mathrel{\raise0.35ex\hbox{$\scriptstyle <$}\kern-0.6em\lower0.40ex\hbox{{$\scriptstyle \sim$}}}}

\def\Wm2{\,\hbox{W}\,\hbox{m}^{-2}}
\def\gsim{\mathrel{\raise0.35ex\hbox{$\scriptstyle >$}\kern-0.6em\lower0.40ex\hbox{{$\scriptstyle \sim$}}}}
\def\lsim{\mathrel{\raise0.35ex\hbox{$\scriptstyle <$}\kern-0.6em\lower0.40ex\hbox{{$\scriptstyle \sim$}}}}
\def\ltsima{$\; \buildrel < \over \sim \;$}
\def\simlt{\lower.5ex\hbox{\ltsima}}
\def\gtsima{$\; \buildrel > \over \sim \;$}
\def\simgt{\lower.5ex\hbox{\gtsima}}

\def\CO12-32{\mathrel{\rm ^{12}CO(3-2)}}
\def\CO12-54{\mathrel{\rm ^{12}CO(5-4)}}
\def\CO12-76{\mathrel{\rm ^{12}CO(7-6)}}
\def\CO13-32{\mathrel{\rm ^{13}CO(3-2)}}
\def\CO13-54{\mathrel{\rm ^{13}CO(5-4)}}
\def\CO13-76{\mathrel{\rm ^{13}CO(7-6)}}
\def\C18O32{\mathrel{\rm C^{18}O(3-2)}}
\def\C18O54{\mathrel{\rm C^{18}O(5-4)}}
\def\C18O76{\mathrel{\rm C^{18}O(7-6)}}

\begin{document}

\title[{\rm $^{13}$CO} and {\rm C$^{18}$O} in a $z$\,=\,2.3
star-forming galaxy]{$^{13}$CO and C$^{18}$O emission from a dense gas
  disk at $z$\,=\,2.3: abundance variations, cosmic rays and the
  initial conditions for star formation}

\author[Danielson et al.]
{\parbox[h]{\textwidth}{
A.\,L.\,R.\ Danielson$^{\,1,* }$,
A.\,M.\ Swinbank$^{\,1}$,
Ian Smail$^{\,1}$,
E.\ Bayet$^{\,2}$,
Paul P.\ van der Werf$^{\,3}$,
P.\ Cox$^{\,4}$,
A.\,C.\ Edge$^{\,1}$,
C.\ Henkel$^{5,6}$ and
R.\,J.\ Ivison$^{\,7,8}$
}
\vspace*{6pt}\\
$^1$Institute for Computational Cosmology, Department of Physics, Durham University, South Road, Durham DH1 3LE, UK\\
$^2$Sub-Department of Astrophysics, University of Oxford, Denys
Wilkinson Building, Keble Road, Oxford, OX1 3RH\\ 
$^3$Leiden Observatory, Leiden University, P.O. Box 9513, NL-2300 RA Leiden, The Netherlands\\
$^4$Institut de Radio Astronomie Millimetrique, 300 rue de la Piscine,
Domaine Universitaire, 38406 Saint Martin d'Heres, France\\
$^5$Max-Planck-Institut fur Radioastronomie, Auf dem Hugel 69, 53121,
Bonn, Germany \\
$^6$Department of Astronomy, King Abdulaziz University,
 P.O. Box 80203, Jeddah, Saudi Arabia\\
$^7$UK Astronomy Technology Centre, Royal Observatory, Blackford Hill, Edinburgh, EH9 3HJ, UK\\
$^8$Institute for Astronomy, University of Edinburgh, Edinburgh, EH9 3HJ, UK\\
$^*$Email: a.l.r.danielson@durham.ac.uk\\
}

\maketitle
\begin{abstract}
  We analyse the spectral line energy distributions (SLEDs) of
  $^{13}$CO and C$^{18}$O for the $J$\,=\,1$\rightarrow$0 up to
  $J$\,=\,7$\rightarrow$6 transitions in the gravitationally lensed
  ultraluminous infrared galaxy SMM\,J2135-0102 at $z$\,=\,2.3. This
  is the first detection of $^{13}$CO and C$^{18}$O in a high-redshift
  star-forming galaxy. These data comprise observations of six
  transitions taken with PdBI and we combine these with $\sim$33\,GHz
  JVLA data and our previous spatially resolved $^{12}$CO and
  continuum emission information to better constrain the properties of
  the inter-stellar medium (ISM) within this system.  We study both
  the velocity-integrated and kinematically decomposed properties of
  the galaxy and coupled with a large velocity gradient model we find
  that the star-forming regions in the system vary in their cold gas
  properties, in particular in their chemical abundance ratios. We
  find strong C$^{18}$O emission both in the velocity-integrated
  emission and in the two kinematic components at the periphery of the
  system, where the C$^{18}$O line flux is equivalent to or higher
  than the $^{13}$CO. We derive an average velocity-integrated flux
  ratio of $^{13}$CO/C$^{18}$O\,$\sim$\,1 which suggests an abundance
  ratio of [$^{13}$CO]/[C$^{18}$O] which is at least 7$\times$ lower
  than that in the Milky Way. This is suggestive of enhanced C$^{18}$O
  abundance, perhaps indicating star formation preferentially biased
  to high-mass stars.  We estimate the relative contribution to the
  ISM heating from cosmic rays and UV of
  (30--3300)$\times10^{-25}$\,erg\,s$^{-1}$ and
  $45\times10^{-25}$\,erg\,s$^{-1}$ per H$_{2}$ molecule respectively
  and find them to be comparable to the total cooling rate of
  (0.8--20)$\times10^{-25}$\,erg\,s$^{-1}$ from the CO. However, our LVG
  models indicate high ($>$100\,K) temperatures and densities
  ($>10^{3}$)\,cm$^{-3}$ in the ISM which may suggest that cosmic rays
  play a more important role than UV heating in this system. If cosmic
  rays dominate the heating of the ISM, the increased temperature in
  the star forming regions may favour the formation of massive stars
  and so explain the enhanced C$^{18}$O abundance. This is a
  potentially important result for a system which may evolve into a
  local elliptical galaxy.

\end{abstract} 


\section{Introduction}

Cold molecular gas in the inter-stellar medium (ISM) in galaxies
provides the raw materials from which stars can form. The physical
properties of the gas determine the range of initial conditions for
star formation.  In local galaxies the ISM typically exhibits a
considerable range of properties, with densities ranging between
10$^{0-7}$\,cm$^{-3}$ and temperatures of $\sim$10\,--\,10000\,K. This
material is heated by a variety of sources, e.g.  the stellar
radiation field, turbulence and cosmic rays, and subsequently cools
through atomic and molecular line emission. Star formation itself
takes place in giant molecular clouds (GMCs) where the gas is at its
densest ($n$(H$_2$)$>$10$^{4}$\,cm$^{-3}$; \citealt{Gao04b};
\citealt{Bergin07}) and the temperatures are typically low
(10--20\,K), making the denser ISM phases the more important to study
when it comes to defining the star-formation process.

The most abundant molecule in giant molecular clouds (GMCs) is
Hydrogen (H$_2$), however, since it lacks a permanent dipole and has
very low mass, it is very challenging to directly observe, as very
high temperatures are required to excite the quadrupolar rotational
transitions. This means that cold, star-forming H$_2$ gas is almost
impossible to directly observe in emission.  The next most abundant
molecule is Carbon Monoxide ($^{12}$C$^{16}$O, hereafter $^{12}$CO),
which has very strong emission lines from pure rotational transitions
at millimetre wavelengths. Since the formation of molecular species
such as $^{12}$CO occurs under similar conditions to H$_2$ formation,
$^{12}$CO emission is commonly used as an H$_2$ tracer species and,
with some assumptions, can be used to derive the mass of H$_2$ likely
to be present in the ISM. The lowest $^{12}$CO rotational transitions
($J$\,=\,1$\rightarrow$0 and 2$\rightarrow$1, hereafter $J_{\rm up}=1$
and $J_{\rm up}=2$) are those typically used for tracing the bulk of
the cold gas in galaxies (e.g. \citealt{Young91}). However, the high
optical depth of lines of the main isotopologue,
$^{12}$CO ($\tau\sim5$\,--\,10 for $J_{\rm up}=1$), at relatively low
column densities, means that in externally-heated clouds, $^{12}$CO
emission is dominated by the warm cloud surfaces, thus limiting the
information that can be deduced about the physical properties of the
high-density molecular gas.  This high optical depth forces the
use of an empirical conversion factor (calibrated from local late-type
galaxies) in order to deduce cold gas mass from $^{12}$CO $J_{\rm
  up}=1$ line emission (so-called $X_{\rm CO}$ or $\alpha_{\rm CO}$
factor; see e.g. \citealt{Dickman86}).  This factor can vary
significantly depending on the average physical and kinematical state
of the cold molecular gas (i.e. UV photodissociation rate, gas
density, column density, kinetic temperature, metallicity; see
\citealt{Bryant96}). The conversion factor $\alpha_{\rm CO}$ (between gas mass
and $^{12}$CO line luminosity, $\alpha_{\rm CO}=M_{\rm H_2}$/L$'_{^{12}\rm
  CO(1-0)}$) has a value of $\sim4.6$ in the disk of the Milky Way,
whereas in local Ultra Luminous Infrared Galaxies (ULIRGs) a
significantly lower average value of $\sim0.8$ is generally preferred,
though with considerable uncertainty (e.g.  \citealt{Downes98};
\citealt{Stark08}; \citealt{PPP12b}).

A more reliable probe of the physical conditions of the cold gas
requires measurements of optically thin tracers, capable of probing
all of the column. $^{13}$CO and C$^{18}$O (rare isotopologues of CO)
can provide this due to their significantly lower abundances (and
hence lower optical depths; $\tau<1$) than $^{12}$CO. $^{13}$C is a
``secondary'' species produced in longer-lived, low-to-intermediate
mass stars, compared to the ``primary'' nature of $^{12}$C which is
formed in and ejected from high-mass stars (\citealt{Wilson94}). Thus,
$^{13}$CO is generally associated with the later stages of star
formation. Less is known about the origin of the optically thin
molecule C$^{18}$O, however, it is generally associated with high-mass
star formation and $^{18}$O is potentially a major constituent of the
winds of massive stars (i.e. \citealt{Henkel93};
\citealt{Prantzos96}). Given the lower abundance of these isotopes
(i.e. [$^{12}$C]/[$^{13}$C] $\sim$\,20--140 \citealt{Martin10};
[$^{16}$O]/[$^{18}$O] $\sim$\,150--200 \citealt{Henkel93}), $^{13}$CO
is expected to be optically thin in all but the highest density,
highest extinction, star-forming cores. In the densest cores C$^{18}$O
may provide a better tracer of H$_2$ due to its lower abundance
and optical depth than $^{13}$CO. In the Milky Way the abundance ratio
of [$^{13}$CO]/[C$^{18}$O]\,$\sim7-8$ (i.e. \citealt{Henkel93}).

Whilst these isotopologues may provide more robust measurements of the
star-forming gas, their low abundances mean that they are very
difficult to detect at high redshift. Indeed, the only high-redshift
detection of $^{13}$CO and C$^{18}$O is from the Cloverleaf quasar
(\citealt{Henkel10}). However, if detections can be made in
star-forming galaxies, then their strengths and line ratios with
respect to other isotopologues will provide a diagnostic of the
physical conditions of the ISM.  For example, GMCs experiencing strong
recent star formation may display elevated C$^{18}$O and $^{12}$CO
abundances relative to $^{13}$CO, in particular in systems
preferentially forming massive stars (i.e. \citealt{Henkel93};
\citealt{Meier04}).

Recently, we discovered a bright, lensed sub-mm galaxy at $z\sim2.3$,
SMM\,J2135$-$0102 (hereafter SMM\,J2135, \citealt{Swinbank10Nature};
\citealt{Ivison10eyelash}; \citealt{Danielson11}). This galaxy is
amplified by a foreground cluster producing three images of the
background galaxy, of which the two brightest are seen adjacent to
each other reflected across the critical curve. These two images have
a combined amplification of $37.5\times$, resulting in an apparent
870\,$\mu$m flux of 106\,$\pm$\,3\,mJy.  Hence, intrinsically the
galaxy has an unlensed 870\,$\mu$m flux of $\sim$3\,mJy and a
far-infrared luminosity of $\sim$2.3$\times$10$^{12}$\,L$_{\odot}$
equivalent to a star formation rate (SFR) of
$\sim$\,400\,M$_{\odot}$\,yr$^{-1}$ (comparable to the local ULIRG
Arp\,220).

\cite{Danielson11} obtain very high signal-to-noise (S/N) detections
of 11 transitions from three molecular and atomic species ($^{12}$CO,
[C{\sc i}] and HCN) and limits on a further 20 transitions from nine
species in SMM\,J2135.  The $^{12}$CO line profiles show multiple
kinematic components with different excitation
temperatures. \cite{Swinbank11} probe the high-resolution kinematics
of the system, identifying four dense star-forming clumps (with
physical scales of $\sim$100--200\,pc) which closely correspond to the
kinematic components. These clumps are embedded within a rotationally
supported disk, $\sim5$kpc in diameter. Moreover, \cite{Danielson11}
show that the cold molecular gas associated with the star formation
appears to be exposed to UV radiation fields 10$^{4}\times$ more
intense than in the Milky Way and suggest that photon-heating should
dominate in the source. However, there have been various studies which
have suggested that in galaxies with high SFR densities (and hence
high supernova rates), heating due to cosmic rays may play a very
significant role in the heating of the H$_{2}$ gas
(e.g. \citealt{Goldsmith78}; \citealt{HD08}; \citealt{Bradford03};
\citealt{Bayet11a}; \citealt{Bayet11b}).  In particular, \cite{PPP10c}
demonstrates the importance of cosmic ray heating over photon heating
in ULIRGs with high densities and high SFRs. In these systems cosmic
rays are capable of penetrating the dense star-forming gas clumps and
volumetrically heating the gas, resulting in kinetic temperatures
8--16$\times$ higher than in UV-shielded star-forming cores in the ISM
of ULIRGs, which in turn can alter the conditions for star formation
and thus the characteristic mass of stars.

In this paper we extend our previous study of SMM\,J2135 from the
$^{12}$CO spectral line energy distribution (SLED) analysis to include
key $J_{\rm up}$\,=1--7, $^{13}$CO and C$^{18}$O line emission, using
the Plateau de Bure Interferometer (PdBI) and the Karl. G Jansky Very
Large Array\footnote{The National Radio Astronomy Observatory is a
  facility of the National Science Foundation operated under
  cooperative agreement by Associated Universities, Inc.} (JVLA).  We
use these data to improve the large velocity gradient (LVG) modelling
of this system, and hence to determine the likely physical conditions
of the cold molecular gas in the ISM.  We compare the $^{13}$CO and
C$^{18}$O line profiles with the optically-thick $^{12}$CO to build a
better understanding of the distribution, properties and kinematics of
the cold molecular gas in this system. We also estimate the likely
heating contributions from cosmic rays and photons in order to
determine the dominant heating process.

In \S\ref{sec:obs} we describe our observations of the molecular
emission from SMM\,J2135. Our observational analysis and results are
described in \S\ref{sec:anal} where we first consider the
velocity-integrated (hereafter `integrated') properties of the system,
followed by decomposing the source into previously identified
kinematic components and deriving their individual cold molecular gas
properties.  In \S\ref{sec:LVG}, we then use large velocity gradient
models to further probe the properties of the cold molecular gas. We
combine the observational and theoretical findings in
\S\ref{sec:discuss} and discuss the possible physical structure and
properties of SMM\,J2135, as well as considering the dominant heating
mechanisms in the source. We give our conclusions in \S
\ref{sec:conc}.  Throughout the paper we use a $\Lambda$CDM cosmology
with H$_0=72$\,km\,s$^{-1}$\,Mpc$^{-1}$, $\Omega_m=0.27$ and
$\Omega_{\Lambda}=1-\Omega_m$ (\citealt{Spergel04};
\citealt{Spergel07}).  We apply a lensing
amplification correction of a factor of $37.5\pm4.5$ to any
luminosities throughout (see \citealt{Swinbank11} for a summary of the
gravitational lensing model).

\section{Observations and Reduction}
\label{sec:obs}

\subsection{Plateau de Bure Interferometer Observations}

\begin{table*}
\caption{Line fluxes and flux ratios for integrated and decomposed spectra}
\input{line_ratios_combined.tex}

\label{tab:lineratio}
\flushleft{\footnotesize{ $^a$We quote 3-$\sigma$ limits for all lines
    which are not formally detected.\\ $^b$Uncertainties on fluxes
    include measurement errors but do not include the flux calibration
    uncertainties, which we estimate as $\sim10$\% at 100\,GHz,
    $\sim15$\% at 165\,GHz and $\sim20$\% at 232\,GHz.\\ $^c$The
    fluxes in Jy\,km\,s$^{-1}$ are observed values.\\
  \item $^d$Fluxes are taken from \cite{Danielson11}, shown here for
    comparison. Flux calibration uncertainties are estimated to
    contribute an additional $\sim5$\% for 30--200GHz and $\sim10$\%
    for 200-300GHz.\\
$^e$The S/N of this line appears to be lower than in the line maps in
Fig.~\ref{fig:all_spec} since the flux is determined by integrating
over a FWZI of 900\,km\,s$^{-1}$ for all lines and the $^{13}$CO(5-4)
line is significantly narrower than the other lines.}}
\end{table*}

We used the six-element IRAM Plateau de Bure
Interferometer\footnote{Based on observations carried out with the
  IRAM Plateau de Bure Interferometer under programme u0b6. IRAM is
  supported by INSU/CNRS (France), MPG (Germany) and IGN (Spain).}
(PdBI) with the {\sc widex} correlator to observe the $J_{\rm up}$=3,
5 and 7 transitions of $^{13}$CO and C$^{18}$O, and the continuum at
$\sim$100, 165 and 232\,GHz respectively. {\sc widex} is a
dual-polarisation correlator with a large bandwidth of 3.6\,GHz and a
fixed frequency resolution of 2\,MHz. Observations were made in the
lowest resolution D-configuration between March 25th 2011 and October
15th 2011.  The frequency coverage was tuned to the systemic redshift
determined from the $^{12}$CO $J$\,=\,1$\rightarrow$0 (hereafter
$^{12}$CO(1--0)) discovery spectrum ($z$\,=\,2.32591;
\citealt{Swinbank10Nature}). We integrated until we reached a noise
level of $\sim$\,0.6\,mJy per 40\,km\,s$^{-1}$ channel in the 100\,GHz
and 165\,GHz observations and $\sim0.8$\,mJy per 50\,km\,s$^{-1}$ in
the 232\,GHz data. The overall flux scale for each observing epoch was
set by MWC\,349 for a majority of the observations, with additional
observations of 2134+004 for phase and amplitude calibrations, due to
the presence of a strong radio recombination line in the emission from
MWC\,349 at 232\,GHz. The data were calibrated, mapped and analyzed
using the {\sc gildas}\footnote{http://www.iram.fr/IRAMFR/GILDAS/}
software package. During the mapping process natural weighting was
applied.  For natural weighting, beam sizes and position angles were
determined to be 4\farcs0\,$\times$\,2\farcs5 and
P.A.\,=\,17.0$^{\circ}$, 3\farcs4\,$\times$\,2\farcs8 and
P.A.\,=\,171.0$^{\circ}$, and 3\farcs4\,$\times$\,2\farcs1 and
P.A.\,=\,2.3$^{\circ}$, for the 100\,GHz, 165\,GHz and 232\,GHz data
respectively.

\begin{figure*}
\centerline{
\psfig{figure=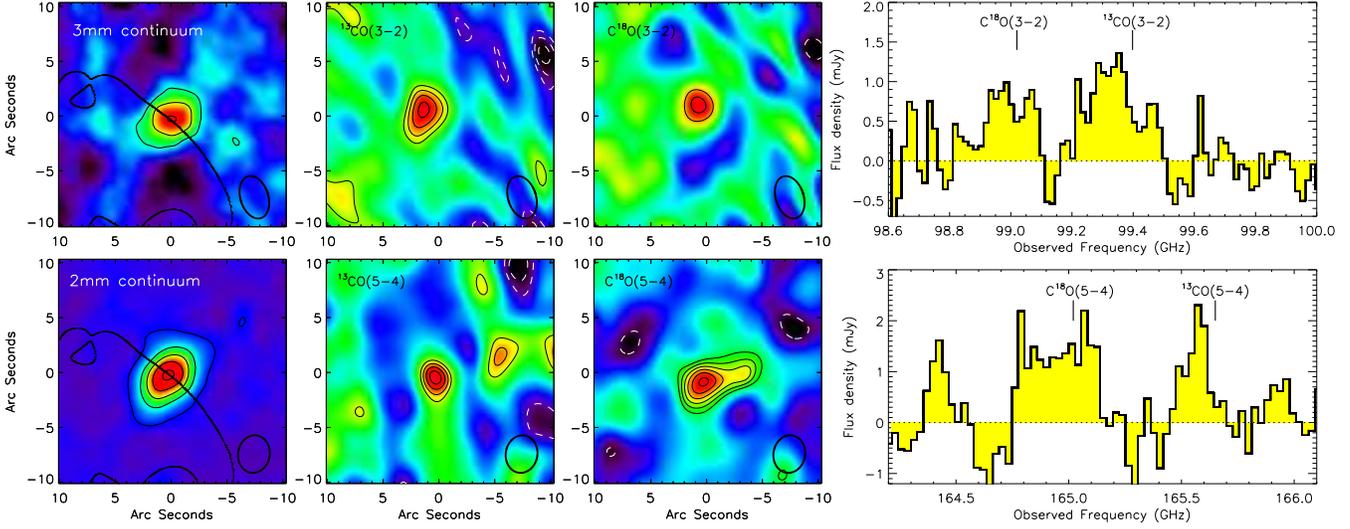,width=7in}}
\caption{{\it Left and centre}: Signal-to-noise maps of the continuum
  and the transitions of $J_{\rm up}=3$ and $J_{\rm up}=5$ $^{13}$CO
  and C$^{18}$O line emission, where the latter two are extracted over
  the frequency extent of the emission lines. The contours on the
  continuum maps represent 5, 10, 15$\sigma$ for the 100\,GHz (3\,mm)
  data and 5, 25, 45$\sigma$ for the 165\,GHz (2\,mm) data. The black
  curve through the continuum maps represents the critical curve from
  the gravitational lensing model \citep{Swinbank11}.  The contours on
  the line maps represent 3, 4, 5, 6$\sigma$ etc. The line maps have
  been continuum-subtracted and then slightly smoothed for display
  purposes. These maps demonstrate that we have significant detections
  of the continuum and the $J_{\rm up}=3$ and $J_{\rm up}=5$
  transitions in SMM\,J2135 and furthermore that the structure and
  extent of the emission varies between the different
  transitions. {\it Right}: The continuum-subtracted spectra (output
  from {\sc gildas}) for the $J_{\rm up}=3$ and $J_{\rm up}=5$
  transitions. The expected central frequencies of the transitions for
  the heliocentric redshift of $z$\,=\,2.32591 are labelled with solid
  lines. The fluxes given in Table~1 have been determined by
  integrating the spectra in the velocity range of $-$350 to
  +550\,km\,s$^{-1}$ (the FWZI of the $^{12}$CO lines in
  \protect\citealt{Danielson11}). The channel widths are
  50\,km\,s$^{-1}$ for both spectra shown.}
\label{fig:all_spec}
\end{figure*}

We show in Fig.~\ref{fig:all_spec} the S/N maps for the $^{13}$CO and
C$^{18}$O observations. The continuum maps are produced by measuring
the median over the off-line continuum in every spatial pixel and
dividing it by the standard deviation of the continuum in the same
frequency range.  To create the line maps, we fit the continuum with a
low order polynomial in every spatial pixel and subtract it from the
cube, we then produce a collapsed image over the full width at zero
intensity (FWZI) frequency range of the line (see below). We then produce S/N maps
from this by dividing the signal by the product of the standard
deviation of the continuum and the square root of the number of
frequency channels in the line. In Fig.~\ref{fig:indiv} we show the
spectra of $^{13}$CO and C$^{18}$O. The 1-D spectra are produced by
integrating within an aperture around the full extent of the emission
region in the cleaned cube. We note that the S/N maps and spectra of
$^{13}$CO and C$^{18}$O shown in Fig.~\ref{fig:all_spec} and
Fig.~\ref{fig:indiv} are continuum-subtracted.

Inspection of the velocity-integrated data cubes
(Fig. \ref{fig:all_spec}) shows strong detections (S/N$\sim$4--8) for
$^{13}$CO(3--2), $^{13}$CO(5--4), C$^{18}$O(3--2) and C$^{18}$O(5--4),
although we only detect continuum at 232\,GHz (i.e. no $^{13}$CO(7--6)
or C$^{18}$O(7--6) detections).  Since the spectra are clearly highly
structured (see also \citealt{Danielson11}) we do not fit a single
Gaussian to determine the integrated flux but instead we integrate the
spectra in the velocity range of $-350$ to +550\,km\,s$^{-1}$ (the
FWZI of the $^{12}$CO composite line in \citealt{Danielson11}; see
Fig.~\ref{fig:composite}). We note that since some of the line
profiles are narrower than others, the S/N of the integrated flux
value may appear low due to integrating over the same FWZI for all
lines which may not be appropriate for i.e. the narrow $^{13}$CO(5--4)
line. For the undetected lines of $^{13}$CO(7--6) and C$^{18}$O(7--6)
we determine a 3\,$\sigma$ upper limit on the fluxes by assuming a
linewidth typical of the average linewidth of the detected lines for
the same species (900\,km\,s$^{-1}$). In Table~1 we provide
line fluxes (with associated measurement uncertainties) or limits for
the various species and we include the $^{12}$CO data from
\cite{Danielson11} for completeness. We estimate additional flux
calibration uncertainties of 10\% at 100\,GHz, 15\% at 165\,GHz and
20\% at 232\,GHz, which are taken into account in all analyses on the
spectral lines but not included in the tabulated flux and error
values.

\subsection{JVLA Observations}

SMM\,J2135 was observed with the JVLA in D-configuration between
September and December 2011. The Ka receiver was tuned to 33.13\,GHz,
covering both the $^{13}$CO(1--0) and C$^{18}$O(1--0) lines.  Standard
amplitude, phase, and bandpass calibration procedures were used. While
we wanted to set up our observations with a scan-averaging time of 1
second, NRAO staff instructed us to use a scan-averaging time of 3
seconds. A bug in the observing software resulted in all the
integration time after the first second of every scan being
discarded. Thus two thirds of the data were lost. As a result the JVLA
spectrum is shallow (containing only 5 hours of integration time) and
only provides upper limits on the $^{13}$CO(1--0) and C$^{18}$O(1--0)
emission. However, we still use these upper limits as constraints for
the modelling in \S\ref{sec:LVG}.

%
%

\section{Analysis and Results}
\label{sec:anal}

We have observed four transitions from $^{13}$CO and C$^{18}$O and
detect both these species in their $J_{\rm up}=3$ and $J_{\rm up}=5$
transition and place sensitive limits on their $J_{\rm up}=1$ and
$J_{\rm up}=7$ emission. We combine this information with the previous
work of \cite{Danielson11} to give a combined dataset with 11
detections and four limits on $^{12}$CO, $^{13}$CO and C$^{18}$O. This
is a unique and unprecedented dataset for a high-redshift galaxy and
allows us to probe the densities, temperatures and chemical abundances
within the ISM of this starburst galaxy.  

\begin{figure*}
\centerline{
  \psfig{figure=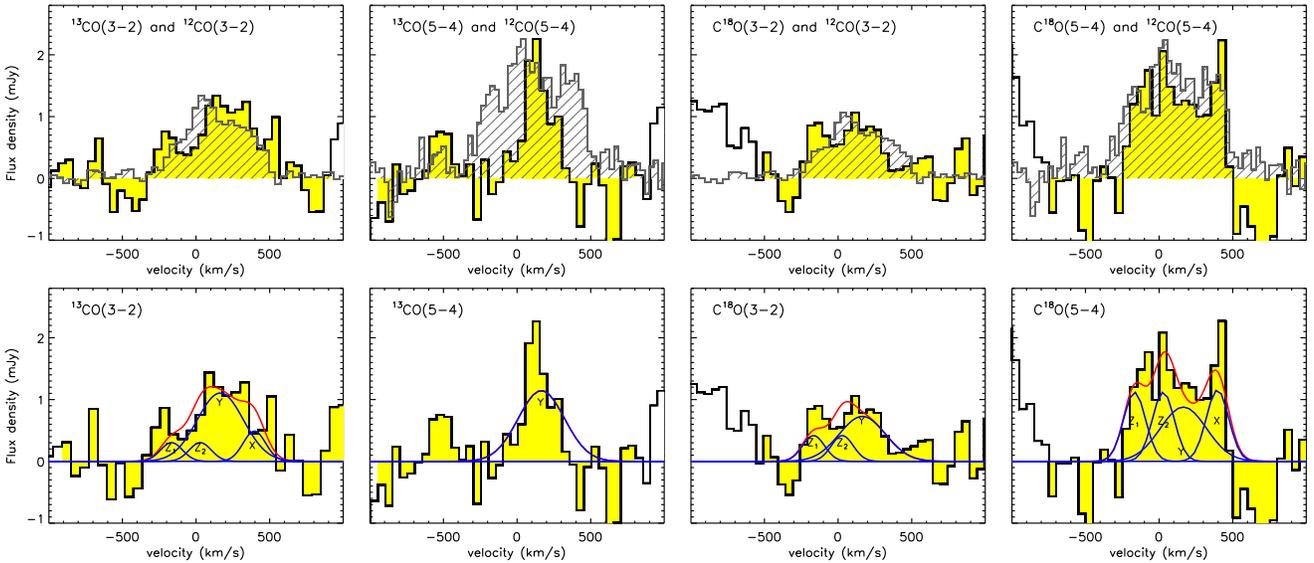,width=7in}}
\caption{{\it Top row}: Continuum-subtracted spectra for the
  individual transitions (filled histograms) plotted to compare the
  line profiles of the $^{13}$CO and C$^{18}$O. Overlaid on each is
  the spectrum for the respective $^{12}$CO transition (normalised by
  peak flux) to compare the velocity structure (hatched region). {\it
    Bottom row}: The spectra for all lines, with the three-component
  kinematic model with components $X$, $Y$, $Z$, taken from the
  Danielson et al. (2011) fit to each line (see \S~\ref{sec:dec_rat}
  and Fig.~\ref{fig:composite}). The channel widths are
  50\,km\,s$^{-1}$ for the $^{13}$CO and C$^{18}$O but
  40\,km\,s$^{-1}$ and 30\,km\,s$^{-1}$ for the $^{12}$CO(3--2) and
  (5--4) respectively. In each case, the continuum was fitted with a
  low-order polynomial and subtracted from the one-dimensional
  spectrum output by {\sc gildas}. It is clear that the line profiles
  differ strongly between the different species. This is likely to be
  due to a combination of factors, potentially including differences
  in optical depth and abundance. The three-component kinematic fit to
  the spectra demonstrates that the different transitions are
  dominated by different kinematic components, i.e. only the
  $Y$-component is required to fit the $^{13}$CO(5--4) line. }
\label{fig:indiv}
\end{figure*}

To compare the line profiles between the isotopologues of $^{12}$CO
and $^{13}$CO, in Fig.~\ref{fig:indiv} we overlay the corresponding
$^{12}$CO transition on the $^{13}$CO spectra, normalised by their
peak fluxes. It is clear that the different species exhibit very
different line profiles and that there are multiple velocity
components detected through $^{13}$CO and C$^{18}$O. These differences
are particularly prominent in $J_{\rm up}=5$ in both $^{13}$CO and
C$^{18}$O where $^{13}$CO(5--4) is significantly narrower than both
the C$^{18}$O(5--4) and the $^{12}$CO lines. Furthermore, as
Fig.~\ref{fig:all_spec} shows, C$^{18}$O(5--4) appears to be spatially
extended along the same direction as the high-resolution $^{12}$CO
maps (\citealt{Swinbank11}) unlike the other lines which are more
compact in their spatial distribution.

%

Considering the unusual profiles, large integrated flux and spatial
extent of the C$^{18}$O(5--4) we test the validity of these data by
splitting the sample into two independent datasets of half the
exposure time and measure the integrated flux in each half of the
data. The S/N in each half of the data is obviously lower than that of
the total sample but we find the C$^{18}$O(5--4) fluxes to be
$0.82\pm0.20$ and $0.83\pm0.25$ which are consistent with each other
and with the total flux of $0.87\pm0.16$. Furthermore, we find that
both the spatial extent and the line profile persist in both samples.

\begin{figure}
\centerline{
  \psfig{figure=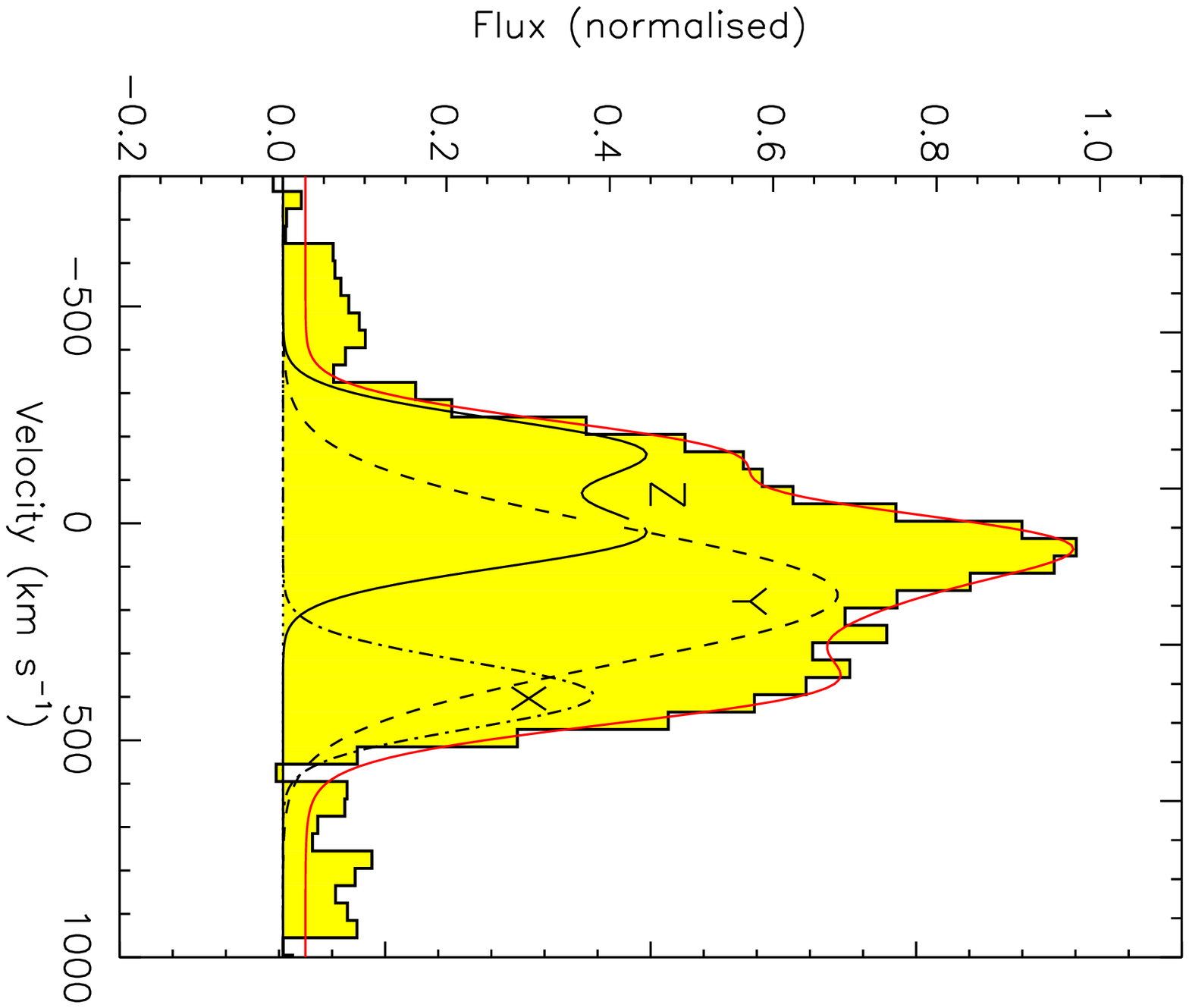,angle=90,width=3.5in}}
\caption{A composite spectrum of the $^{12}$CO spectra of SMM\,J2135
  from $J_{\rm up}=1-8$, normalised by the peak flux, taken from
  \protect\cite{Danielson11}. This composite is used to define a FWZI
  range over which to sum the total flux ($-$350 to
  550\,km\,s$^{-1}$). Furthermore, utilising the high signal-to-noise
  of the $^{12}$CO emission lines, this composite spectrum is used (in
  \protect\citealt{Danielson11}) to fit and determine the
  three-component ($X$, $Y$ and the double-peaked $Z$) kinematic model
  plus continuum that we apply to our $^{13}$CO and C$^{18}$O emission
  lines in this work.}
\label{fig:composite}
\end{figure}

Figs.~\ref{fig:indiv} and \ref{fig:lineratio} clearly show the
presence of multiple kinematic components with different $^{13}$CO and
C$^{18}$O line strengths. In \cite{Danielson11} we kinematically
decompose the $^{12}$CO emission line spectra in SMM\,J2135 into
multiple velocity components (i.e. Fig.~\ref{fig:composite}). In
\S\ref{sec:dec_rat} we fit this same model to the $^{13}$CO and
C$^{18}$O lines and discuss the kinematically decomposed properties of
SMM\,J2135. However, before attempting to disentangle the internal
variations of the ISM conditions within this system, we analyse the
integrated properties of the galaxy and attempt to draw some broad
conclusions, providing a comparison to our subsequent
kinematically-resolved analysis.

\subsection{Integrated Properties}

\subsubsection{Integrated Line Ratios}
\label{sec:lineratio_int}
\begin{figure*}
\centerline{ \psfig{figure=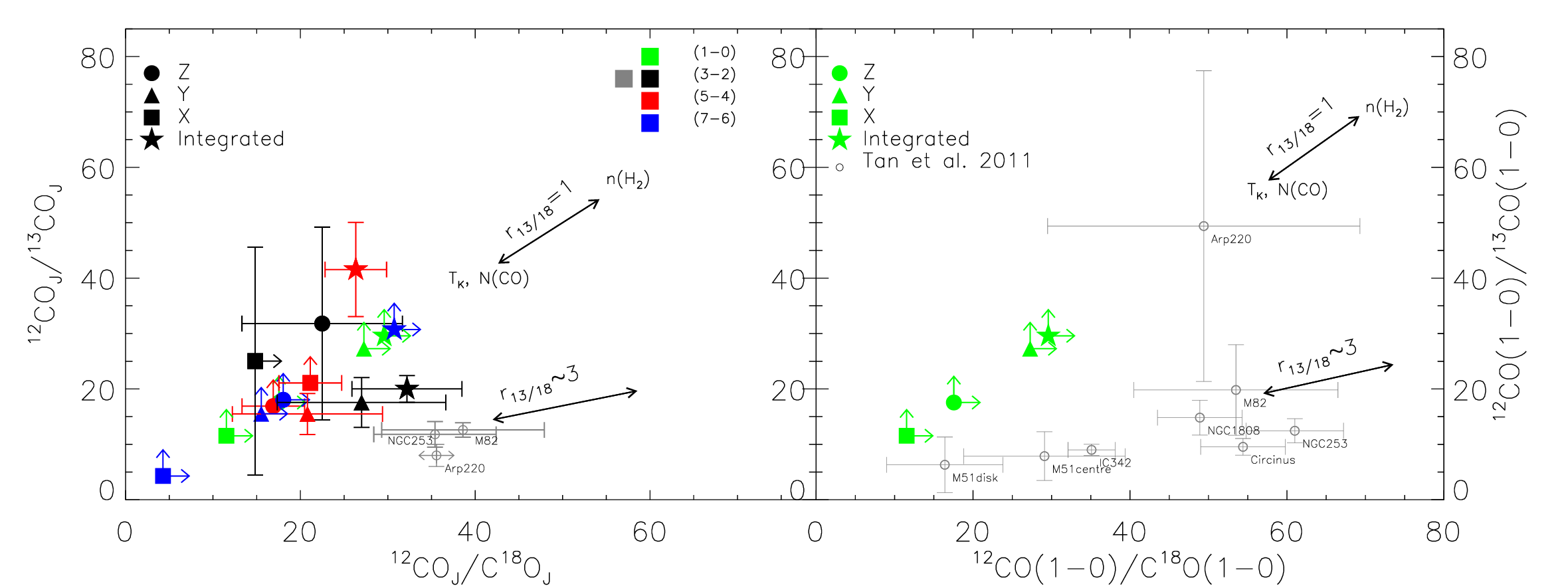,width=7in}}
\caption{{\it Left:} Flux ratios of $^{12}$CO/$^{13}$CO versus
  $^{12}$CO/C$^{18}$O for the integrated emission from SMM\,J2135 for
  the different transitions available. We also plot the individual
  components of the three-component kinematic model.  For comparison
  we show the flux ratios of the star-forming systems Arp\,220
  (\protect\citealt{Greve09}), M\,82 (\protect\citealt{Petitpas00})
  and NGC\,253 (\protect\citealt{Harrison99}). The literature values
  are ratios using the $J_{\rm up}=3$ transition in order to compare
  directly to our data, however, C$^{18}$O(3--2) fluxes are only
  published for M\,82 and NGC\,253, therefore for Arp\,220,
  $^{12}$CO(3--2)/C$^{18}$O(3--2) is determined by multiplying the
  known ratio in $J_{\rm up}=1$ by the ratio of
  ($^{12}$CO(3--2)/C$^{18}$O(3--2))/($^{12}$CO(1--0)/C$^{18}$O(1--0))\,$\sim0.72$
  for M\,82. Since the estimate for $^{12}$CO(3--2)/C$^{18}$O(3--2) in
  Arp\,220 is a derived value we do not give a horizontal error bar
  but note that there is large uncertainty in this estimate. Our
  integrated $^{12}$CO/C$^{18}$O flux ratio is consistent with that
  for M\,82, whereas our $^{12}$CO/$^{13}$CO is higher than that for
  M\,82. The vectors demonstrate the effect on the line ratios of
  varying the temperature ($T_{\rm K}$), density ($n$(H$_{2}$)) and
  column density ($N(^{12}$CO)) in the LVG models described in
  \S\ref{sec:LVG} whilst fixing the other parameters in the models at
  their best-fit values. Increasing the
  temperature and column density causes a decrease in the line ratios,
  whereas, increasing the density causes an increase in the line
  ratios. $r_{13/18}$ is defined as $N(^{13}$CO)/$N($C$^{18}$CO) and we
  include a vector for a lower column density ratio to demonstrate the
  effect of decreasing the abundance ratio (see \S\ref{sec:int_sleds}
  for details). The different kinematic components appear to fit with
  different ratios of $r_{13/18}=1-10$, implying that the ISM
  conditions may vary between the components. {\it Right:} We also
  show our $3\sigma$ lower limits from JVLA for the (1--0) emission in
  order to compare to a broader range of literature data
  (\protect\citealt{Tan11}). Our data lies at higher
  $^{12}$CO(1--0)/$^{13}$CO(1--0) than local bright infrared galaxies,
  with the lower limits on $^{12}$CO(1--0)/C$^{18}$O(1--0) being lower
  than the lower limits of the local galaxy ratios, implying
  potentially lower than average $^{13}$CO and higher than average
  C$^{18}$O (assuming $^{12}$CO is optically thick in all cases). This
  could be a genuine abundance effect.  }
\label{fig:lineratio}
\end{figure*}

Fig.~\ref{fig:lineratio} shows the integrated flux ratios of
$^{12}$CO/$^{13}$CO and $^{12}$CO/C$^{18}$O for SMM\,J2135. These flux
ratios can be used as estimates of the abundance ratios but only if
the molecules have the same excitation properties and all lines are
optically thin (both of which- as we see later- are unlikely to be
true). We compare the integrated flux ratios in SMM\,J2135 with those
of star-forming galaxies and ULIRGs in the nearby Universe. Given that
line ratios vary with the transition observed, in order to be
consistent, we must compare ratios in $J_{\rm up}=3$ with literature
ratios in $J_{\rm up}=3$. Published measurements of extragalactic
C$^{18}$O ($J_{\rm up}$\,$>1$) emission are rare and for star-forming
galaxies we only have this information for the local starburst galaxy
M\,82 which \cite{Petitpas00} mapped in $J_{\rm up}=3$ for $^{12}$CO,
$^{13}$CO and C$^{18}$O, from which we quote the line ratios from the
central position in the galaxy, and for the central $23''$ of the
starburst NGC\,253 from \cite{Harrison99}. In Fig.~\ref{fig:lineratio} we
also include the measured line ratios in $J_{\rm up}=5$ and $J_{\rm
  up}=7$ in SMM\,J2135, for completeness. There is a much more
extensive galaxy sample available for $^{13}$CO and C$^{18}$O in the
$J_{\rm up}=1$ transition, therefore we also use our limits on $J_{\rm
  up}=1$ in order to compare our data more directly to other galaxies.
For the literature line ratios of $^{12}$CO(1--0)/$^{13}$CO(1--0) and
$^{12}$CO(1--0)/C$^{18}$O(1--0) we use \cite{Tan11}.

Our $^{12}$CO/$^{13}$CO integrated flux ratios for SMM\,J2135 are
$\sim20\pm2$ and $>31$ in $J_{\rm up}=3$ and $J_{\rm up}=1$
respectively, both of which are higher than local star-forming
galaxies and ULIRGs. For example, in M\,82 (Fig.~\ref{fig:lineratio})
lower values of $^{12}$CO(3--2)/$^{13}$CO(3--2)\,=\,$12.6\pm1.5$
(\citealt{Petitpas00}) and
$^{12}$CO(1--0)/$^{13}$CO(1--0)\,=\,$19.8\pm8.17$ (\citealt{Tan11})
are found. Furthermore, our integrated $^{12}$CO(1--0)/$^{13}$CO(1--0)
ratio lies above that for local infrared luminous sources from
\cite{Tan11}. Only Arp\,220 has a comparably large value of
$^{12}$CO(1--0)/$^{13}$CO(1--0)\,$\sim$\,$50\pm30$. 

There are many potential explanations for the observed elevated
$^{12}$CO/$^{13}$CO flux ratios: {\it i}) Turbulence due to mergers or
winds can cause the $^{12}$CO line to broaden, thus decreasing the
optical depth ($\tau$) of the line and increasing the
$^{12}$CO/$^{13}$CO intensity ratio without the need for a change in
the relative abundances of the isotopologues \citep{Aalto95}; {\it
  ii}) In UV radiation-dominated photodissociation regions on the
surfaces of molecular clouds, $^{12}$CO, with its higher optical
depth, more effectively self-shields than $^{13}$CO (and C$^{18}$O),
thus $^{13}$CO is more easily photodissociated than $^{12}$CO,
resulting in an elevated $^{12}$CO/$^{13}$CO ratio. Similarly,
C$^{18}$O is more easily photodissociated than $^{13}$CO resulting in
a higher $^{13}$CO/C$^{18}$O ratio \citep{Bally82}; {\it iii}) As
$^{13}$C is a secondary product from a later stage of nuclear
processing than $^{12}$C, age can play a role in determining the
$^{12}$CO/$^{13}$CO ratio, such that in systems with many newly
forming stars $^{12}$CO/$^{13}$CO can be much higher than in older
systems where there has been enough time to synthesise $^{13}$C
(\citealt{Henkel10}).  Thus, there is a trend of increasing
$^{12}$CO/$^{13}$CO with decreasing metallicity, i.e. ULIRGs, which
are characterised by lower metallicity, typically exhibit higher
$^{12}$C/$^{13}$C ratios (\citealt{Casoli92};
\citealt{Henkel93a,Henkel98}; \citealt{Meier04} \citealt{Genzel12});
{\it iv}) \cite{Casoli92b} find that the $^{12}$CO/$^{13}$CO can be
enhanced in regions of recent bursts of star formation, particularly
of massive stars, as $^{12}$C is overproduced by nucleosynthesis
relative to $^{13}$C leading to an overabundance of $^{12}$CO; {\it
  v}) Similarly, infall of unprocessed gas from the disk into the
nuclear region(s) of starbursts may lead to enhanced
$^{12}$CO/$^{13}$CO ratios; {\it vi}) Finally, chemical fractionation
may also affect the observed $^{12}$CO/$^{13}$CO ratio, i.e.
\begin{equation}
^{13}\rm C^{+}+ {^{12}\rm CO} \rightarrow {^{12}C^{+}} + {^{13}CO} +\Delta E_{35K},
\end{equation}
\citep{Watson76}, which enhances $^{13}$CO relative to $^{12}$CO
leading to a lower value for $^{12}$CO/$^{13}$CO.

In contrast to the elevated $^{12}$CO/$^{13}$CO flux ratio, as
Fig.~\ref{fig:lineratio} shows, SMM\,J2135 appears to have a lower
$^{12}$CO/C$^{18}$O flux ratio than local infrared luminous
galaxies. This implies either a deficiency in $^{12}$CO or, more
likely, enhanced C$^{18}$O. However, there are a few local
  systems which display similar characteristics. For example,
  \cite{Meier01} find very low $^{12}$CO/C$^{18}$O integrated
  intensity ratios in the centre of the star-forming galaxy IC\,342
  ($^{12}$CO(1--0)/C$^{18}$O(1--0)\,=\,24$\pm7$ in the central trough
  compared to $>110$ in the off-arm regions), suggesting a very high
  abundance of C$^{18}$O relative to $^{12}$CO in the star-forming
  regions.  The mechanism for $^{18}$O synthesis is not well
understood, however, a high abundance of C$^{18}$O is thought to be
due to the enrichment of C$^{18}$O from $^{18}$O rich, massive star
ejecta and winds (i.e. \citealt{Henkel93}; \citealt{Meier01}). Indeed,
since $^{18}$O is a secondary product produced in massive stars during
He-burning by $^{14}$N$^{18}$O, an overabundance of C$^{18}$O in
systems which are preferentially producing massive stars may be
expected (e.g. due to initial mass function (IMF) variations;
\citealt{Henkel93}). However, it is interesting to note that initial
nitrogen abundances and therefore metallicities would have to be
almost solar to facilitate the production of such large quantities of
$^{18}$O, which would be unusual at $z=2.3$\footnote{Using the
  mass-metallicity relation for $z\sim1.5$ galaxies from Stott et al.
  (in prep) we estimate a metallicity for SMM\,J2135 of
  $\sim0.3Z_{\odot}$}.

It is also interesting to note that \cite{PPP96} find a relatively low
average line ratio of $^{13}$CO/C$^{18}$O\,=\,3.3 in NGC\,1068 for
$J_{\rm up}=1$ and infer that this is either due to an intrinsically
low [$^{13}$CO]/[C$^{18}$O] abundance ratio or optical depth effects
(even though both $^{13}$CO and C$^{18}$O are generally considered to
be optically thin tracers). Similarly, in Arp\,220, the line
  ratio $^{13}$CO(1--0)/C$^{18}$O(1--0) is found to be only
  $1.0\pm0.3$ \citep{Greve09}.

Thus, our integrated flux ratios of
$^{13}$CO(3--2)/C$^{18}$O(3--2)\,$=1.6\pm0.4$ and
$^{13}$CO(5--4)/C$^{18}$O(5--4)\,$=0.44\pm0.18$ in SMM\,J2135 are
unusually low compared to those found in local starbursts and
ULIRGs. This is likely to imply an enhanced abundance of
C$^{18}$O. Fig.~\ref{fig:lineratio} demonstrates the abnormality of
SMM\,J2135, particularly with regard to low $^{12}$CO/C$^{18}$O
ratios, further emphasising the possibility of a high abundance of
C$^{18}$O implied by the unusually low $^{13}$CO/C$^{18}$O line
ratios.

%

\subsection{Kinematically Resolved Properties}
\label{sec:dec_rat}

\subsubsection{Kinematically Resolved Model}
\label{sec:kin_model}
Using high-resolution mapping, \cite{Swinbank11} demonstrated that the
velocity structure in SMM\,J2135 coincides with the clumpy structure
in the disk. Hence the $\sim$100--200~pc star-forming clumps observed
in the Smithsonian Submillimeter Array (SMA) rest-frame 260\,$\mu$m
map coincide with the clumps observed in the cold molecular gas and
these in turn correspond with the kinematic components in the high
resolution $^{12}$CO spectra (\citealt{Swinbank11}).

\begin{table}
\caption{Kinematically decomposed model fit parameters}
\begin{center} 
\smallskip
\begin{tabular}{lll}
\hline
\noalign{\smallskip}
Component & $v$ & $\sigma$    \\
          &  (km\,s$^{-1}$)      &      (km\,s$^{-1}$)      \\ 
\hline
$Z_{2}$ & $-167\pm9$ & $75\pm8$ \\
$Z_{1}$ & $28\pm9$ & $75\pm8$ \\
$Y$ & $165\pm13$ & $157\pm17$ \\
$X$ & $396\pm9$ & $76\pm9$ \\
\hline
\label{tab:decomparam}
\end{tabular}
\end{center}
\footnotesize{ Note: Velocities given with respect to the heliocentric
  redshift of $z=$\,2.32591}
\end{table}

To decompose the $^{13}$CO and C$^{18}$O lines into multiple kinematic
components, we adopt the kinematic model derived from
\cite{Danielson11}. This comprises three Gaussian components: $X$, $Y$
and $Z$, where the latter is a coupled double Gaussian. This model
provides a reasonable approximation of the observed line profiles
(Fig.~\ref{fig:composite}). To be consistent with the $^{12}$CO
analysis, we fix the central velocities and linewidths of the
components $X$, $Y$ and $Z$ to the values derived using $^{12}$CO but
allow the intensities to vary when fitting to the new observations
(see Table~2). However, as in \cite{Danielson11} the intensities of
the double Gaussian describing component $Z$ are tied to be equal. We
show this three-component model in Fig.~\ref{fig:composite}. The
central velocities of the components are 396\,km\,s$^{-1}$,
165\,km\,s$^{-1}$, $-70$\,km\,s$^{-1}$ for the $X$, $Y$ and $Z$
components respectively.  The best fit models for $^{13}$CO and
C$^{18}$O are overlaid in Fig.~\ref{fig:indiv} and fluxes (and
associated errors) derived from the individual components are listed
in Table~1.  The $^{13}$CO and C$^{18}$O lines are much fainter
(typically $\sim20-40\times$ lower) than the $^{12}$CO and hence the
detections have lower S/N than the previous $^{12}$CO
decomposition. As a result, there are degeneracies in fitting the
three-component model to some of the lines, however, this approach
provides a useful means of consistently comparing the data with our
previous $^{12}$CO decomposition.  Some of the components are
undetected in certain transitions and we have strong variations in the
line ratios between components (see Table~1 and
Fig.~\ref{fig:lineratio}) and in \S\ref{sec:LVG} we attempt to model
these. 

There is a moderate amplification gradient across the source, causing
the $X$-component fluxes to appear fainter than the
$Z$-component. This is demonstrated in \cite{Swinbank11} where the
amplification increases across the source from $\sim20-50$ from the
$X$-component to $Z$-component. However, our kinematic decomposition
is not influenced by the differential magnification as the velocities
map uniquely to spatial positions in the lens plane and thus to a
single magnification factor.

%
%
\begin{figure}
\centerline{
\psfig{figure=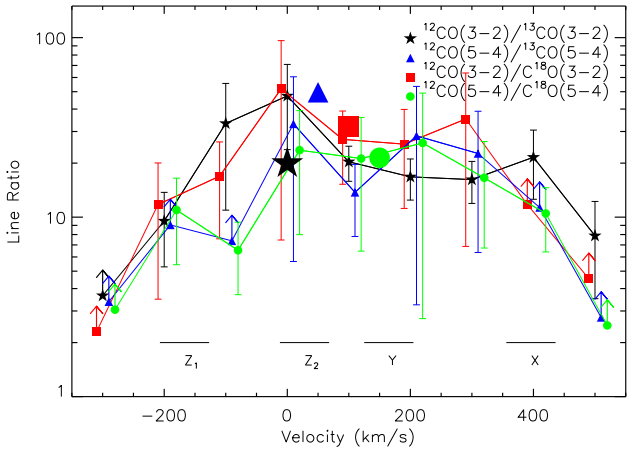,width=3.6in}}
\caption{The variation of the estimated line ratios as a function of
  velocity across the spectral lines of $^{12}$CO, $^{13}$CO and
  C$^{18}$O in the $J_{\rm up}=3$ and 5 transitions. We have re-binned
  the spectra into channels of 100\,km\,s$^{-1}$ and offset the data
  sets by 10\,km\,s$^{-1}$ from each other for clarity. The errors are
  determined from the standard deviation of the off-line
  100\,km\,s$^{-1}$ channels. Lower limits at the edges are due to the
  $^{13}$CO and C$^{18}$O having a narrower FWZI than the
  $^{12}$CO. These are 3\,$\sigma$ lower limits. We indicate the
  velocity centres of the main kinematic components.  Overall there is
  no strong trend with velocity. The integrated line ratios over the
  FWZI of 900\,km\,s$^{-1}$ of each line are shown as larger filled
  symbols placed arbitrarily at 0, 50, 100 and 150\,km\,s$^{-1}$. They
  are broadly consistent with each other for all transistions. }
\label{fig:opt_depth}
\end{figure}

It is clear from Fig.~\ref{fig:indiv} that the line profiles of
$^{12}$CO, $^{13}$CO and C$^{18}$O display strong variation in the
$^{12}$CO/$^{13}$CO and $^{12}$CO/C$^{18}$O with velocity, possibly
tracing variation in the optical depth of $^{13}$CO and C$^{18}$O or
abundance variations.  To directly compare the lines, we re-bin the
$^{12}$CO, $^{13}$CO and C$^{18}$O spectra in J$_{\rm up}$\,=\,3 and 5
to 100\,km\,s$^{-1}$ channels and in Fig.~\ref{fig:opt_depth} we plot
the variation of these line ratios with velocity over the
$\sim900$\,km\,s$^{-1}$ FWZI of the $^{12}$CO. We calculate the ratio
of the flux density in each channel ($S_{\nu}(12)/S_{\nu}(i)$), where
S$_{\nu}$ is the flux density in mJy and {\it i} refers to either
$^{13}$CO or C$^{18}$O.  In Fig.~\ref{fig:opt_depth} we show
horizontal bars indicating the central velocity of the individual
kinematic components.  We see little measurable difference between the
line ratios as a function of velocity however, there is an overall
trend that the line ratios are higher towards the kinematic centre
($Y$) of the source than at the edges.

As shown by Fig.~\ref{fig:lineratio} (see also Table~1) the
$^{12}$CO$/^{13}$CO, $^{12}$CO$/$C$^{18}$O and $^{13}$CO/C$^{18}$O
line ratios of the individual kinematic components ($X$, $Y$, $Z$) in
$J_{\rm up}=3$ and 5 are all within 1$\sigma$ of the galaxy-integrated
values, aside from the $^{12}$CO(5--4)/$^{13}$CO(5--4) and
$^{13}$CO(5--4)/C$^{18}$O(5--4) which are significantly
different. This is mainly due to the $^{13}$CO(5--4) line being
significantly narrower than the other lines and hence in
Fig.~\ref{fig:indiv} only the $Y$-component is required to fit the
line.

In Fig.~\ref{fig:lineratio}, we also show vectors which demonstrate
the effect of varying the abundance ratio of
$N(^{13}$CO)/$N$(C$^{18}$O) ($r_{13}/r_{18}$; see
\S\ref{sec:int_sleds} for details). It appears that the integrated
source, $X$- and $Z$-components follow the trend of the vector
corresponding to $N(^{13}$CO)\,/\,$N($C$^{18}$O)\,$=1$, whereas the
different transitions of the $Y$ component are better fit with a higher
abundance closer to $N(^{13}$CO)\,/\,$N($C$^{18}$O)\,$\sim3$ which
may imply that the ISM conditions vary between the components.

In all components we find up to $\sim$10$\times$ lower
$^{13}$CO/C$^{18}$O flux ratios than the typical value of
[$^{13}$CO]/[C$^{18}$O]$\sim$\,4 (\citealt{Wang04}) found in local
star-forming galaxies. This is similar to Arp\,220, which exhibits low
line ratios of $^{13}$CO/C$^{18}$O\,$\sim1$ at its centre in both
$J_{\rm up}$\,=\,1 and 2 (\citealt{Matsushita09}; \citealt{Greve09}),
attributed to a high abundance of C$^{18}$O arising from a recent
starburst. As noted earlier, lower than expected abundance ratios may
result from $^{18}$O enriched gas being ejected from high mass
stars. Thus it may imply a bias towards high mass star formation. The
$^{12}$C/$^{13}$C abundance ratio is not affected as much since
$^{13}$C arises from later stages of lower-mass star formation.

\section{LVG Modelling}
\label{sec:LVG}
Whilst the comparison of line ratios of different isotopologues is
useful, the degeneracies between density, temperature, optical depth
and abundance mean that to better constrain the physical properties of
the ISM we need to model the spectral line energy distributions
(SLEDs). In this section we investigate both the galaxy-integrated
properties and the kinematically-resolved component properties to
better understand the ISM in this system.

\subsection{Model Description}
In \cite{Danielson11} we analysed the $^{12}$CO SLED to attempt to
constrain the physical conditions of the molecular gas in SMM\,J2135,
through large velocity gradient (LVG) modelling
(\citealt{Weiss05b}). Here we can include in our analysis the lower
abundance $^{13}$CO and C$^{18}$O SLEDs in order to better determine
the ISM conditions. We therefore use the non-LTE radiative transfer
code, {\sc radex}, developed by \cite{Vandertak07}, to model the SLEDs
and determine the likely physical conditions within SMM\,J2135 using
our complete $^{12}$CO, $^{13}$CO and C$^{18}$O SLEDs. We first apply
this analysis to the integrated SLEDs, but then motivated by the
strong line ratio differences seen within this source
(\S\ref{sec:dec_rat}; Figs.~\ref{fig:indiv}, \ref{fig:lineratio} and
\ref{fig:opt_depth}) we apply the same approach to the kinematically
decomposed SLEDs.

Briefly, {\sc radex} solves the radiative transfer equations by
assuming an isothermal and homogeneous medium without large-scale
velocity fields and assuming a certain geometry to describe the photon
escape probabilities. Three different geometries are available but we
choose the model mimicking a uniform expanding sphere (LVG
approximation), giving a corresponding escape probability formalism of
$\beta_{LVG}=(1-e^{-\tau})/\tau$ (e.g. \citealt{Goldreich74}). In the
model, the equations of statistical equilibrium are iteratively
solved, beginning by assuming low $\tau$ (optically thin) for all
emission lines. To derive the physical parameters using this model,
molecular collisional rates are required. We used collisional rates
from the Leiden Atomic and Molecular Database
(LAMDA)\footnote{http://www.strw.leidenuniv.nl/$\sim$moldata/} as
recommended in \citet{Vandertak07}. We assume H$_{2}$-CO collisional
excitation rates from \cite{Flower01} and a cosmic microwave
background (CMB) temperature of $\sim9$\,K (2.73\,K redshifted to
$z=2.3$).

The main input variables in the LVG model we use are: the column
density $N(X)$ of the species ($X$) considered ($^{12}$CO, $^{13}$CO
or C$^{18}$O) in cm$^{-2}$, the line width (FWHM) $\Delta V$ for each
transition in km\,s$^{-1}$, the molecular hydrogen volume gas density
$n$(H$_{2}$) in cm$^{-3}$, the kinetic temperature $T_{\rm K}$ in K,
and the abundance ratios (i.e. $^{12}$C$/^{13}$C). These variables
must either be fixed or fitted for and are described below. In the
following we describe our method for deriving the best model
parameters from our LVG model.

\subsection{Input Parameter Selection}
\label{sec:input}
In order to select a suitable velocity gradient ($\Delta V$) we
require constraints on the size and structure of the source. From
previous work (\citealt{Swinbank10Nature}, 2011;
\citealt{Danielson11}), SMM\,J2135 appears to comprise at least four
100--200\,pc star-forming clumps embedded in a rotating gas disk with
a diameter of $\sim$5\,kpc.  We initially take the average full width
at half maximum (FWHM) of the disk emission ($\sim$500\,km\,s$^{-1}$)
as an estimate of the average velocity field that the gas in the
system is experiencing ($\Delta V$). The model outputs
velocity-integrated line intensities by multiplying the
model-estimated radiation temperature by $1.06\times\Delta V$
(assuming a Gaussian profile). This initial method is crude given that
it assumes uniform excitation conditions throughout the entire source,
which given the clear structure in the line profiles is unphysical,
however, it is useful to first determine the average molecular gas
properties.  For each of the species $^{12}$CO, $^{13}$CO and
C$^{18}$O, we carry out simulations and produce model grids with {\sc
  radex} for the following parameter ranges: kinetic temperature
ranging between $T_{\rm K}$=10\,--\,200\,K in 10\,K steps; gas density
ranging between $n$(H$_2$)\,=\,$10^3-10^7$cm$^{-3}$ in 0.5\,dex steps
and molecular line column densities $N(^{12}$CO, $^{13}$CO,
C$^{18}$O)=$10^{10}-10^{20}$cm$^{-2}$ in intervals of 0.5\,dex. See
\cite{Bayet13} for the choice of input parameters and ranges and
for a more complete description of the {\sc radex} model we implement.

\subsection{Model Outputs}
\label{sec:output}
In an ideal situation, where all lines are optically thin, we could
make the assumption that the integrated line intensity ratio of
$^{12}$CO/$^{13}$CO is equivalent to the ratio of the column
densities, $N(^{12}$CO)/$N(^{13}$CO), which is equivalent to the
atomic abundance ratio [$^{12}$C]/[$^{13}$C]. We make this assumption
in our analysis but we note that since $^{12}$CO has a higher optical
depth than $^{13}$CO, the measured $^{12}$CO/$^{13}$CO line intensity
ratio provides only a lower limit on the actual
[$^{12}$CO]/[$^{13}$CO] abundance ratio.  We do not fix
$N(^{12}$CO)/$N(^{13}$CO) when searching for the best-fitting model
but instead we allow it to vary, making it an output of the models in
order to determine the best-fit abundance ratio.  For each unique set
of parameters T$_{\rm K}$, $n$(H$_{2}$) and $N(^{12}$CO, $^{13}$CO,
C$^{18}$O) the model outputs are fluxes in cgs units (which we
  convert to Jy\,km\,s$^{-1}$ in order to compare directly to our
  measured fluxes), excitation temperature T$_{\rm ex}$, brightness
  temperature T$_{\rm b}$ and optical depths for $^{12}$CO, $^{13}$CO
and C$^{18}$O from J$_{\rm up}$\,=\,1--18. In order to also search for
the optimal abundance ratios we construct a grid of models for all
combinations of T, $n$(H$_{2}$), $N(^{12}$CO),
$N(^{12}$CO)/$N(^{13}$CO) and $N(^{13}$CO/$N($C$^{18}$O).  We use
  $N(^{13}$CO)/$N($C$^{18}$O) as a proxy for the
  [$^{13}$CO]/[C$^{18}$O] abundance ratio. To reduce the number of
possible models when searching for the best-fit model we allow only
five different discrete values of $N(^{13}$CO)/$N($C$^{18}$O), ranging
between 0.1--10 (e.g. \citealt{Penzias83}; \citealt{Zhu07}) and
similarly only six discrete values of $N(^{12}$CO)/$N(^{13}$CO),
ranging between $\sim$1--300 in steps of 0.5 dex. Furthermore, we
  restrict the allowed parameter space such that the implied source
  size ($\Omega_{\rm S}$) of the emitting region has a radius of
  $<$5kpc (the diameter of the galaxy is $\sim$5\,kpc as derived from
  high-resolution $^{12}$CO observations from PdBI and JVLA;
  \citealt{Swinbank11}).  We therefore set an upper limit of 5\,kpc
  for the radius of SMM\,J2135:
\begin{equation}
  \Omega_{\rm S}=L'_{\rm CO}/(T_{\rm b} \Delta V D_{\rm A}^2)
\end{equation}
This most noticeably restricts the minimum allowed column density of
the models to N($^{13}$CO)$>10^{15}$cm$^{-2}$.  

\subsection{Integrated Spectral Line Energy Distribution}
\label{sec:int_sleds}

The LVG analysis from \cite{Danielson11} showed that the $^{12}$CO
galaxy-integrated SLED is best-fit by a two-phase model: a `hot',
dense phase ($T_k\sim60$\,K, $n($H$_2)\sim10^{3.6}$cm$^{-3}$) likely
associated with the four dense star-forming clumps and contributing
60\% of the total luminosity over all the $^{12}$CO lines, and a
`cold', diffuse ($T_k\sim25$\,K, $n($H$_2)\sim10^{2.7}$cm$^{-3}$)
phase probably corresponding to an extended gas phase in which the clumps are
embedded. We therefore begin by attempting a two-phase fit to the new data
to test the effect of the additional constraints of $^{13}$CO and
C$^{18}$O SLEDs on the results of the LVG modelling.

\begin{figure*}
\centerline{
\psfig{figure=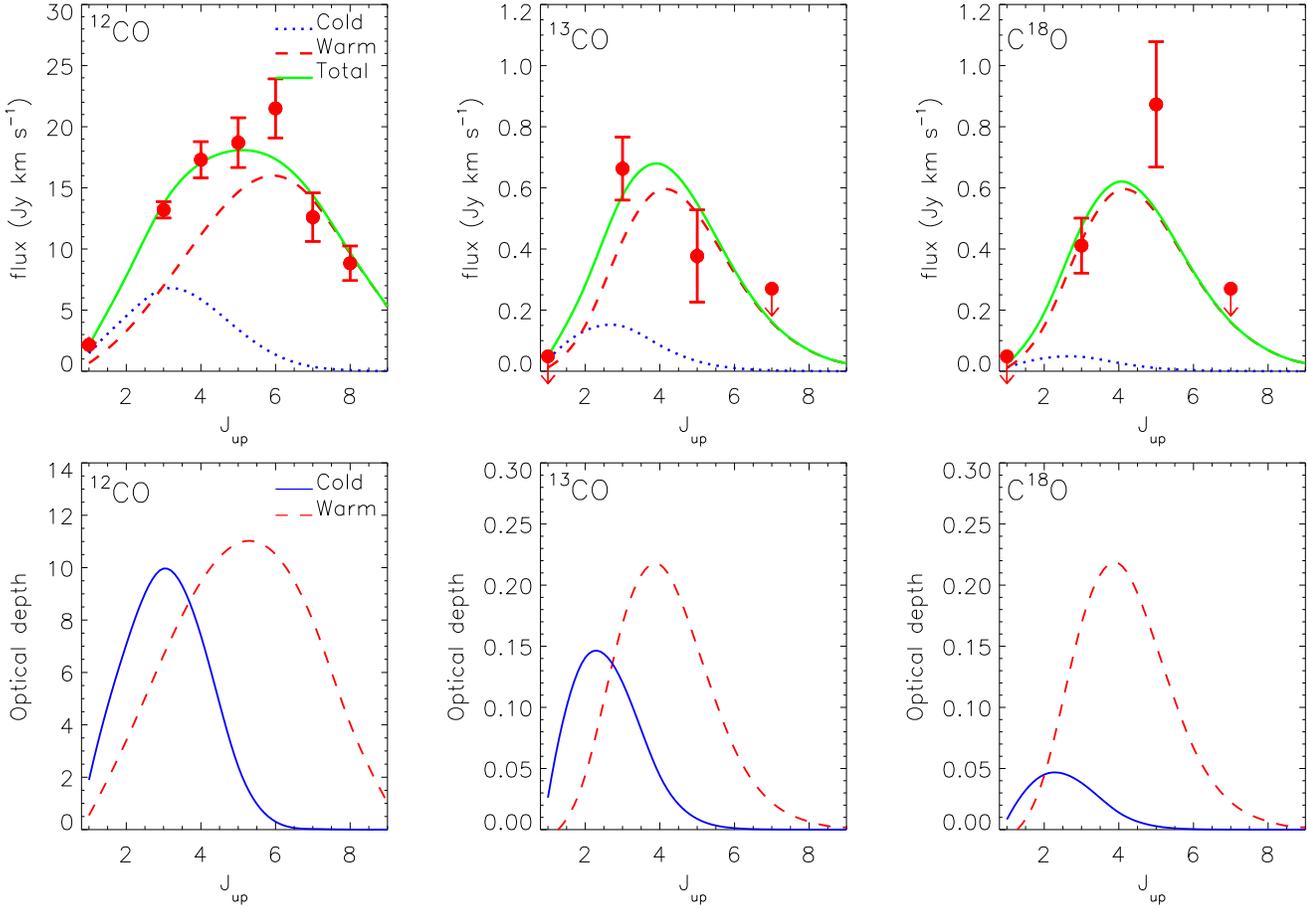,width=8in}}
\caption{ Top: SLEDs for the integrated $^{12}$CO, $^{13}$CO and
  C$^{18}$O emission from SMM\,J2135. The green solid curve is the best
  two-phase LVG fit to the SLEDs for all three species at once
  (described in \S\ref{sec:LVG}). Also shown are the constituent `hot'
  and `cold' phases in red-dashed and blue-solid curves
  respectively. Upper limits shown are at a 2$\sigma$ level for the
  undetected lines.  Bottom: The distribution of the optical depth
  values for the best-fit `cold' and `hot' phases shown as blue-solid
  and red-dashed curves respectively.  }
\label{fig:int_sleds}
\end{figure*}

For this two-phase fit we first split the large model grid into `cold'
and `hot' models with a divide at 50\,K and search the grid for the
optimum combination of `cold' models and `hot' models using a $\chi^2$
calculation. We identify the best two-phase model with the
minimum $\chi^2$ value then find those two-phase models for which the
predicted $^{12}$CO, $^{13}$CO and C$^{18}$O SLEDs are within $\Delta
\chi^{2}=1\sigma$ of the best-fit. In fitting the SLEDs with two phases
simultaneously we have a total of six degrees of freedom per ISM
phase: $n$(H$_{2}$), $T_{\rm K}$, $N$($^{12}$CO),
$N(^{12}$CO)/$N(^{13}$CO), $N(^{13}$CO)/$N($C$^{18}$O) and a
normalisation factor. For 12 degrees of freedom we therefore have a
1\,$\sigma$ confidence limit of $\Delta \chi^{2}=13.7$. Our general
formula for calculating $\chi^2$ is:
\begin{equation}
\chi^2=\sum_{\rm n=1-7}(^{\rm i}\rm
  CO_{J=n}-^{\rm i}CO_{model,J=n})^2/(\alpha_{^{i}\rm CO_{J=n}})^2,
\end{equation} 
where $^{\rm i}$CO represents either $^{12}$CO, $^{13}$CO or
C$^{18}$O, $J_{\rm up }=1, 3, 5, 7$ for $^{13}$CO and C$^{18}$O and
$J_{\rm up }=1, 3, 4, 5, 6, 7, 8$ for $^{12}$CO. We include our upper
limits on $J_{\rm up}=1$ and 7 as constraints in the $\chi^{2}$
fitting, setting the value to ($2\pm1)\sigma$ (in
Figs.~\ref{fig:int_sleds} and \ref{fig:dec_sleds} we show these data
as $2\sigma$ upper limits).  In this equation $\alpha_{^i\rm
  CO_{J=n}}$ represents the error on the line flux constraints.

The measured SLEDs for the three species are shown in
Fig.~\ref{fig:int_sleds} and we show the best-fit combination of `hot'
and `cold' models.
%
%
Table~3 gives the best-fit parameters and range of allowed parameters
for models that lie within $\Delta \chi^{2}=13.7$ of the best-fit. The
best-fit parameters for the `cold' and `hot' phases respectively are:
$\chi_{min-reduced}^2=4.1$,
$N(^{13}$CO)/$N($C$^{18}$O)\,=\,$10^{0.5}$ and $10^{0}$,
$n$(H$_{2}$)\,=\,$10^{3}$cm$^{-3}$ and
$10^{4}$cm$^{-3}$, $N(^{12}$CO)\,/\,$N(^{13}$CO)\,=\,10$^{2}$
for both phases, $N(^{12}$CO)\,=\,$10^{19.5}$cm$^{-2}$ and
$10^{20}$cm$^{-2}$ and $T_{\rm K}$\,=\,50\,K and 90\,K.  The
addition of the constraints of $^{13}$CO and C$^{18}$O seems to
require models with higher temperatures than those found in
\cite{Danielson11} (25\,K and 60\,K) but the densities are similar
($n$(H$_{2}$)\,=\,$10^{2.7-3.6}$cm$^{-3}$). It is important to note
that there is degeneracy in the temperature and density such that it
is possible that an optimal solution in fact lies at higher density
and lower temperature or lower density and higher temperature. To
demonstrate this effect we fix all the model parameters at their
optimised values and determine the effect of varying each parameter in
the models in turn. This is shown by the vectors in
Fig.~\ref{fig:lineratio} ($r_{13}$/$r_{18}$) which show that
increasing the temperature and column density causes a decrease in the
line ratios of $^{12}$CO/$^{13}$CO and $^{12}$CO/C$^{18}$O, whereas,
increasing the density causes an increase in the line ratios.  The
$N(^{13}$CO)/$N($C$^{18}$O) ratios for both the `hot' and `cold'
phases imply that throughout the system (and particularly in the `hot'
phase) there may be an enhanced C$^{18}$O abundance compared to
star-forming galaxies locally.

\begin{figure*}
\centerline{
\psfig{figure=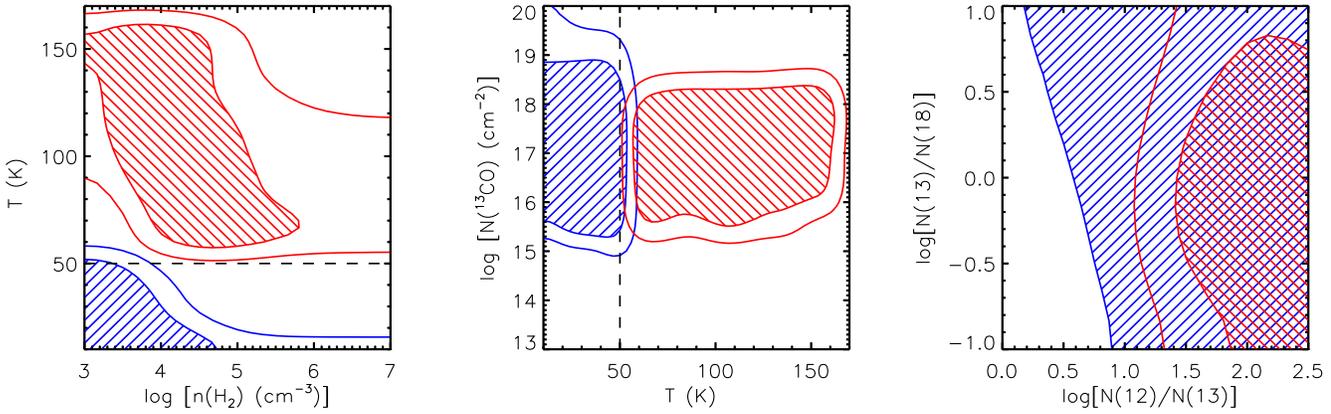,width=8.0in,angle=90}}
\caption{Parameter space distributions of the best-fitting `cold'
  (blue) and `hot' (red) phases for different combinations of
  parameters. The contours represent the models lying within $\Delta
  \chi^{2}=1$ and 2$\sigma$ which for 12 degrees of freedom
  corresponds to models within $\Delta \chi^{2}$ of 13.7 and 21.3
  respectively. The dotted lines on the left and middle plots
  demonstrate that for the two-phase model fits we restricted the
  `cold' phase to have $T\le50$\,K and for the `hot' phase we required
  that it have $T>50$\,K. The hashed regions show the region of
  parameter space occupied by all the models within $\Delta
  \chi^{2}=1\sigma$ of the best-fit model. Since there are a large
  number of degrees of freedom compared to the number of data points
  we have relatively poor constraints on the allowed parameter
  space. This provides further motivation to kinematically decompose
  the emission to test if we can better constrain the parameter space
  of the individual kinematic components. }
\label{fig:int_param}
\end{figure*}

We note that although these are the best-fit values, a large range of
parameters is allowed within $\Delta \chi^{2}=13.7$ of the best-fit
model.  This large range of parameters is demonstrated by
Fig.~\ref{fig:int_param} where we plot likelihood contours
representing the models that lie within $\Delta \chi^2$\,=\,1 and
2$\sigma$ of the best-fit. For the `cold' phase, $n$(H$_2$) is at the
lower limit of the allowed parameter space and for the `hot' phase
$N(^{12}$CO) is at the upper limit of the allowed parameter space,
which means that the best-fitting model may in fact require even
higher $N(^{12}$CO) and lower $n$(H$_2$), making it difficult to make
firm conclusions. However, we note that we initially selected an upper
limit of $N(^{12}$CO)\,=\,10$^{20}$cm$^{-2}$ for our parameter space,
as a higher column density than this would imply extreme values of
$N$(H$_{2}$). We also caution that since the errors on the $^{12}$CO
fluxes are smaller and there are more measured fluxes for $^{12}$CO
than for the less abundant species, we are slightly weighted towards
the $^{12}$CO data when searching for an optimum model. Since in
\cite{Danielson11} we derived lower temperature and densities for the
two phases than we derive here, this may imply that giving the
$^{12}$CO SLEDs more weighting may bias the temperature and density
towards lower values.

%

Since the models we employ also contain optical depth as part of the
output, in Fig.~\ref{fig:int_sleds} we also show the optical depth for
each transition of $^{12}$CO, $^{13}$CO and C$^{18}$O for the best-fit
`hot' and `cold' phases. This demonstrates that $^{12}$CO is
substantially optically thicker ($\sim50\times$) than the $^{13}$CO
and C$^{18}$O, as expected. Furthermore, they demonstrate that the
`hot' phases are all optically thicker than the `cold' phase. This is
particularly prominent for the C$^{18}$O for which the `hot' phase has
a $>$4 times higher optical depth than the `cold' phase, however, they
would both still be classed as optically thin as $\tau\ll 1$.

With the species abundance ratios and temperatures derived from {\sc
  radex} we can now estimate the gas mass of SMM\,J2135 using the
optically thin $^{13}$CO emission and compare this to previous
estimates based on $^{12}$CO in \cite{Danielson11}.

\subsubsection{Gas Mass}
\label{sec:gasmass}
In \cite{Danielson11} we used the $^{12}$CO(1--0) line luminosity and
the $^{12}$CO SLED to constrain a large velocity gradient model which
predicted a total cold molecular gas mass of $M_{\rm
  gas}\sim1.4-4.0\times10^{10}$M$_{\odot}$ for SMM\,J2135. This
suggested that the empirical conversion factor between CO luminosity
and gas mass ($\alpha_{\rm CO}=M_{\rm H_2}$/L$'_{^{12}\rm CO(1-0)}$)
is likely to be $\alpha_{\rm CO}\sim2$ for this system, which is
higher than the canonical $\alpha_{\rm CO}\sim0.8$ normally assumed
for high-redshift ULIRGs. \cite{Swinbank11} use high-resolution
mapping of the $^{12}$CO(6--5) emission to develop a kinematic model
for SMM\,J2135 and estimate a dynamical mass of $M_{\rm
  dyn}=(6.0\pm0.5)\times10^{10}$M$_{\odot}$ within
$\sim5$\,kpc. Together with an estimated stellar mass of
$M_{*}=(3 \pm 1)\times10^{10}$M$_{\odot}$ (\citealt{Swinbank10Nature})
this suggests a cold molecular gas mass of $M_{\rm
  gas}\la4.5\times10^{10}$M$_{\odot}$, which is consistent with our
predictions from the LVG analysis in \cite{Danielson11}.

However, constraints on the total mass of the H$_2$ reservoir can be
derived using measurements of $^{13}$CO and C$^{18}$O, with the
advantage that these lower abundance species are expected to be
optically thin and so able to trace the bulk of the cold molecular gas
including the denser gas typically not probed by $^{12}$CO
emission. Of course, it is still necessary to assume an abundance
ratio between these less abundant isotopologues and H$_{2}$ in order
to use our measured fluxes to determine the cold gas masses.

To estimate the masses we must first determine the intrinsic source
brightness temperature.  The integrated line intensity is $I_{\rm
  CO}$\,=\,$\int T_{\rm mb}dV$ (K\,km\,s$^{-1}$), and is obtained from
the beam-diluted brightness temperature. This must be corrected for
the redshift in order to get the velocity-integrated source brightness
temperature. Following \cite{Solomon97} where the line luminosity is
$L'_{\rm CO}$\,=\,$T_{\rm b} \Delta V \Omega_{\rm S} D_{\rm A}^2$, we
derive a velocity-integrated source brightness temperature of $T_{\rm
  b} \Delta V$\,=\,$L'_{\rm CO}/\Omega_{\rm S} D_{\rm A}^2$, where
$\Omega_{\rm S}$ is the solid angle subtended by the source (in
steradians) and $D_{\rm A}$ is the angular size distance (in
parsecs). The line luminosity can be calculated from our observed
quantities via:
\begin{equation}
L'_{\rm CO}=3.25\times10^7 S_{\rm
    CO} \Delta V \nu_{\rm obs}^{-2}D_{\rm L}^2(1+z)^{-3}/\mu (\rm K km
  s^{-1}pc^2), 
\end{equation}
where $S_{\rm CO} \Delta V$ is the velocity-integrated flux density in
Jy\,km\,s$^{-1}$, $\nu_{\rm obs}$ is the observed frequency of the
transition considered in GHz, $D_{\rm L}$ is the luminosity distance
in Mpc and $\mu$ is the amplification factor due to gravitational
lensing of $\mu=37.5\pm4.5$.

To derive the mass we use:
\begin{equation}
M_{\rm H_2}=N_{\rm H_2}\mu_{\rm G} m_{\rm H_2}\Omega_{\rm S} D_{\rm A}^2,
\end{equation}
where $m_{\rm H_2}$ is the molecular mass of Hydrogen
($3.346\times10^{-27}$kg), $\mu_{\rm G}$ is the mean atomic weight of
the gas (1.36; including the contribution from Helium assuming a 24\%
Helium mass abundance; \citealt{Scoville86}) and $N_{\rm H_2}$ is the
column density in cm$^{-2}$. We calculate $N_{\rm H_2}$ following
\cite{Meier04}:
\begin{equation}
  N(\rm H_2)_{^i\rm CO}=A \times \frac{[\rm H_2]}{[^{12}\rm CO]} \frac{[^{12}\rm CO]}{[^{i}\rm CO]}\frac{e^{^iE_{\rm u}/T_{\rm ex}}}{e^{^iE_{\rm u}/T_{\rm ex}}-1} I_{^{i}\rm CO},
\end{equation}
assuming these species to be optically thin and assuming local thermal
equilibrium (LTE) for all levels. Here $I_{^i\rm CO}$ is the
velocity-integrated source brightness temperature of the chosen
transition in the chosen species {\it i} (in K\,km\,s$^{-1}$);
$^iE_{\rm u}$ is the rotational energy ($h \nu/k$\,=\,5.5\,K for
J$_{\rm up}$\,=\,1); we adopt an average abundance ratio of
[$^{12}$CO]/[H$_2$]$=8.5\times10^{-5}$ (Galactic value;
\citealt{Frerking82}) and $A$ is a constant determined following
\cite{Scoville86}:
\begin{equation}
A=\frac{3k}{8\pi^3B\mu^2}\frac{e^{hBJ_1(J_1+1)/kT_{\rm ex}}}{(J_1+1)}, 
\end{equation}
where $B$ is the rotational constant ($5.5101\times10^{10}$\,s$^{-1}$)
and $\mu$ is the permanent dipole moment ($\mu$\,=\,0.1098 Debyes for
$^{13}$CO where 1
Debye\,=\,$10^{-18}$\,g$^{1/2}$\,cm$^{5/2}$\,s$^{-1}$). Since
$I_{^{i}\rm CO}$\,=\,$T_{\rm b}\Delta V$\,=\,$L'_{\rm CO}/\Omega_{\rm
  S} D_{\rm A}^2$ the mass equation therefore simplifies to:
\begin{equation}
M_{\rm H_2}=A \times \mu_{\rm G} m_{\rm H_2}\frac{[\rm
  H_2]}{[^{12}\rm CO]} \frac{[^{12}\rm CO]}{[^{i}\rm
  CO]}\frac{e^{^iE_{\rm u}/T_{\rm ex}}}{e^{^iE_{\rm u}/T_{\rm
      ex}}-1}L'_{\rm CO},
\end{equation}
where $L'_{\rm CO}$ must be converted into K\,km\,s$^{-1}$\,cm$^2$.

From the LVG modelling we predict best-fit values of the excitation
temperature of $^{13}$CO(3--2) of T$_{\rm ex}$\,=\,11.3\,K (with a
1$\sigma$ range of 9.1--16.8\,K) for the cold component and T$_{\rm
  ex}$\,=\,35.4\,K (with a 1$\sigma$ range of 18.9-155.3\,K) for the
warm component. Our best-fit value of the abundance ratio of
[$^{12}$CO/$^{13}$CO] is $\sim100$ for both the cold and warm
components. This is significantly higher than the values typically
assumed for starburst galaxies of [$^{12}$CO]/[$^{13}$CO]\,=\,40
(\citealt{Henkel93}; \citealt{Wilson94}). However, we note that there
is considerable uncertainty in the abundance ratio within the models
on which the derived mass depends strongly.  Furthermore, there are
strong variations even within galaxies. For example, within the Milky
Way the $^{12}$C/$^{13}$C atomic abundance ratio varies from $\sim$20
in the Galactic centre to $>$100 in the outer parts of the Galaxy
(\citealt{Wilson94}; \citealt{Wouterloot96}). Generally, starburst
galaxies and ULIRGs exhibit a $^{12}$C$/^{13}$C\,$>30$
(e.g. \citealt{Henkel93}; \citealt{Henkel10}). However, in the
starburst galaxy, M\,82, $^{12}$C$/^{13}$C\,$\gsim$140
(\citealt{Martin10}).

If we calculate a mass for the warm and cold components separately,
using our best-fit value for [$^{12}$CO]/[$^{13}$CO], we find $M_{\rm
  H_2}^{warm}$\,=\,(5.4--7.2)$\times10^{9}$M$_{\odot}$ and $M_{\rm
  H_2}^{cold}$\,=\,(2.1--7.8)$\times10^{9}$M$_{\odot}$ giving a range
in total mass of $M_{\rm
  H_2}$\,=\,(0.8--1.5)$\times10^{10}$M$_{\odot}$ with the hot and cold
components contributing approximately equally to the total mass.
However, clearly there is uncertainty in the abundance ratio
used. Indeed, if we take the maximum abundance ratio within 1$\sigma$
of the best fit ([$^{12}$CO]/[$^{13}$CO]$\sim$300) we derive an upper
limit on the best-fit total mass of M$_{\rm
  H_2}\sim2.5\times10^{10}$M$_{\odot}$. The upper limits of this
estimate of H$_{2}$ mass (derived from $^{13}$CO(3--2)) are consistent
with the dynamical limits on M$_{\rm gas}$ derived by $M_{\rm
  dyn}=M_{*}+M_{\rm gas}$, with
$M_{*}=(3\pm1)\times10^{10}$M$_{\odot}$ and $M_{\rm
  dyn}=(6.0\pm0.5)\times10^{10}$M$_{\odot}$ (\citealt{Swinbank11};
neglecting the contributions of dark matter which is not expected to be
dynamically dominant on these scales in a high-redshift galaxy).

Finally we caution that the magnification factor that we apply is a
luminosity-weighted galaxy-integrated factor, however, due to the
critical curve passing through SMM\,J2135, there is a moderate
amplification gradient across the source. This is demonstrated in
\cite{Swinbank11} where the amplification increases across the source
from $\sim20-50$. However, the $^{13}$CO line that we are using to
estimate the gas mass covers the full velocity range of the source and
therefore the integrated magnification we use is probably appropriate.

\subsection{Kinematically Resolved SLEDs}
\label{sec:decsled}

It can be seen from Fig.~\ref{fig:int_sleds} that in both the
$^{12}$CO and C$^{18}$O SLEDs the best-fit model does not peak as high
in J$_{\rm up}$ as the data. In \cite{Danielson11} the model fit
peaked at J$_{\rm up}$\,=\,6 whereas adding in the constraints of
$^{13}$CO and C$^{18}$O results in a poorer fit to the high-excitation
end of the $^{12}$CO SLED. Fundamentally, the problem is that if the
$^{13}$CO and C$^{18}$O emission are both optically thin, then the
difference in their SLED profiles implies that they are not tracing
the same gas and therefore, even two-phase LVG models will struggle to
fit both species simultaneously. Furthermore, while it is possible to
find suitable fits to the data using a two-phase fit, it is difficult
to constrain the allowed parameter space due to the large number of
degrees of freedom. The velocity structure in the lines suggests
multiple components (Figs.~\ref{fig:indiv} and \ref{fig:lineratio})
and high-resolution mapping of SMM\,J2135 demonstrates that the
velocity structure in SMM\,J2135 coincides with the clumpy structure
observed in the disk (\citealt{Swinbank11}). This therefore motivates
us to carry out LVG modelling using the SLEDs derived from our
kinematic decomposition. However, since we are now considering
subcomponents within our system, we must employ a $\Delta V$ according
to the physical scale of the clumps. The gas clumps identified by
\cite{Swinbank11} are $\sim$100--200pc across and have an average FWHM
of $\sim200$km\,s$^{-1}$. However, given that these clumps are likely
to be comprised of much smaller structures, the average velocity that
the gas in the clumps is experiencing is likely to be much lower than
their overall FWHM. We therefore search for models with a $\Delta
V=$\,50\,km\,s$^{-1}$. We include the same constraints as before in
the $\chi^2$ fitting but since we do not have the S/N or enough
datapoints to do a two-phase fit to each kinematic component we use a
single-phase fit to each velocity component with a total of six
degrees of freedom: $n$(H$_{2}$), $T_{\rm K}$, $N$($^{12}$CO),
$N(^{12}$CO)/$N(^{13}$CO), $N(^{13}$CO)/$N($C$^{18}$O) and
normalisation factor. The six degrees of freedom give a 1$\sigma$
confidence limit of $\Delta \chi^{2}=7.0$. Again, if we do not detect
a component in a particular transition, we include upper limits as
constraints, setting the value to (2$\pm1)\sigma$.

\begin{figure*}
\centerline{
\psfig{figure=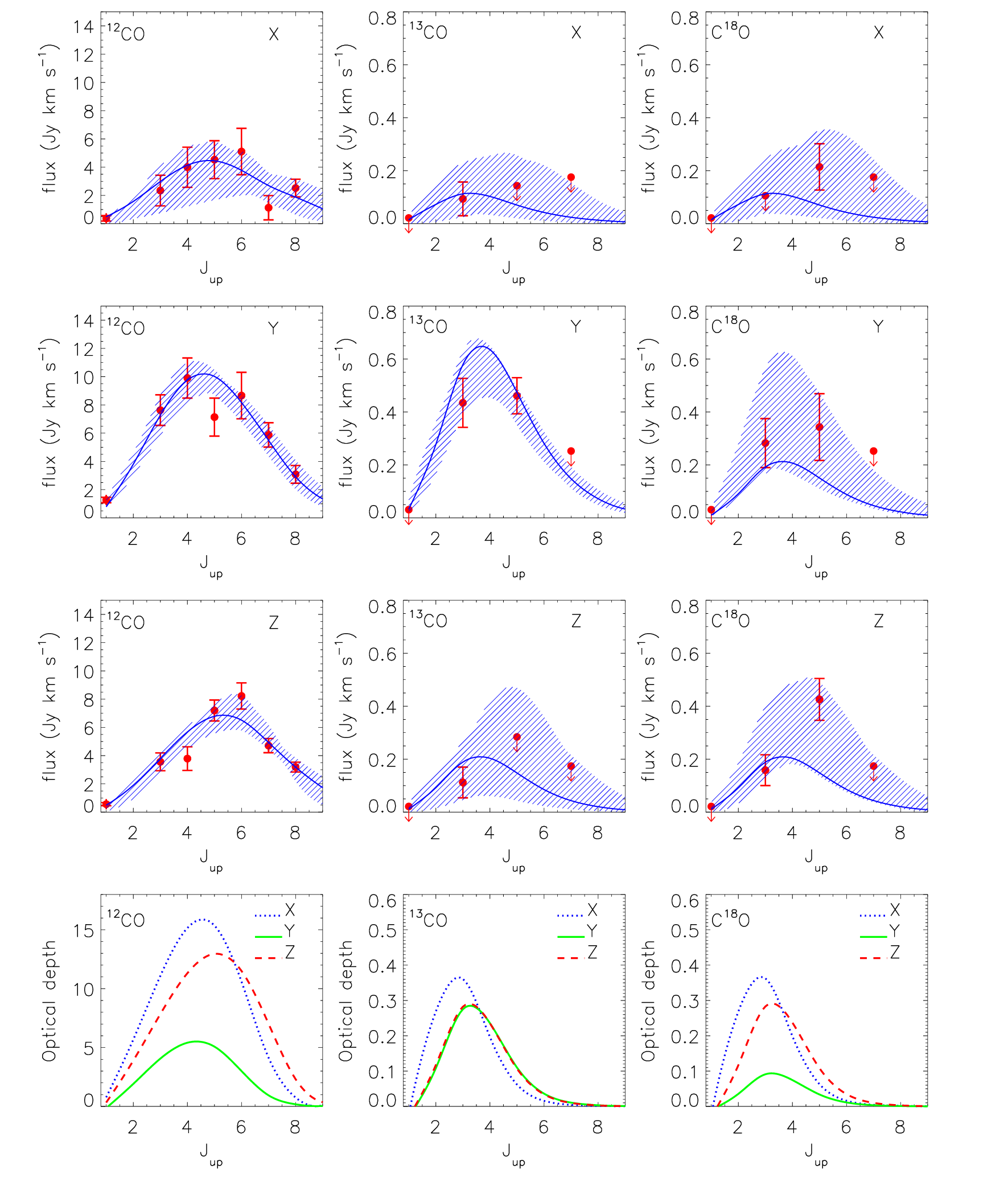,width=7.3in}}
\caption{ SLEDs for the individual kinematic components $X$, $Y$ and
  $Z$ in $^{12}$CO, $^{13}$CO and C$^{18}$O. Overlaid in hashed blue
  is the range in LVG model fits that are within $\Delta
  \chi^2\le1\sigma$ of the best-fit to the $^{13}$CO, C$^{18}$O and
  $^{12}$CO data (all models that lie within $\Delta
  \chi^2\le7.0$). The solid curve represents the model with the
  minimum total $\chi^2$ value for the fit to all the SLEDs for that
  component. Upper limits are plotted at the 2$\sigma$ level. The
  $^{13}$CO and C$^{18}$O SLEDs clearly peak at different J$_{\rm up}$
  but this is not well-described by the models. {\it Bottom:} The
  distribution of the optical depth values for the best-fit models to
  the individual components, demonstrating that the optical depth of
  $^{12}$CO is$\sim40\times$ higher than that of $^{13}$CO and
  C$^{18}$O.}
\label{fig:dec_sleds}
\end{figure*}

\begin{figure*}
\centerline{
\psfig{figure=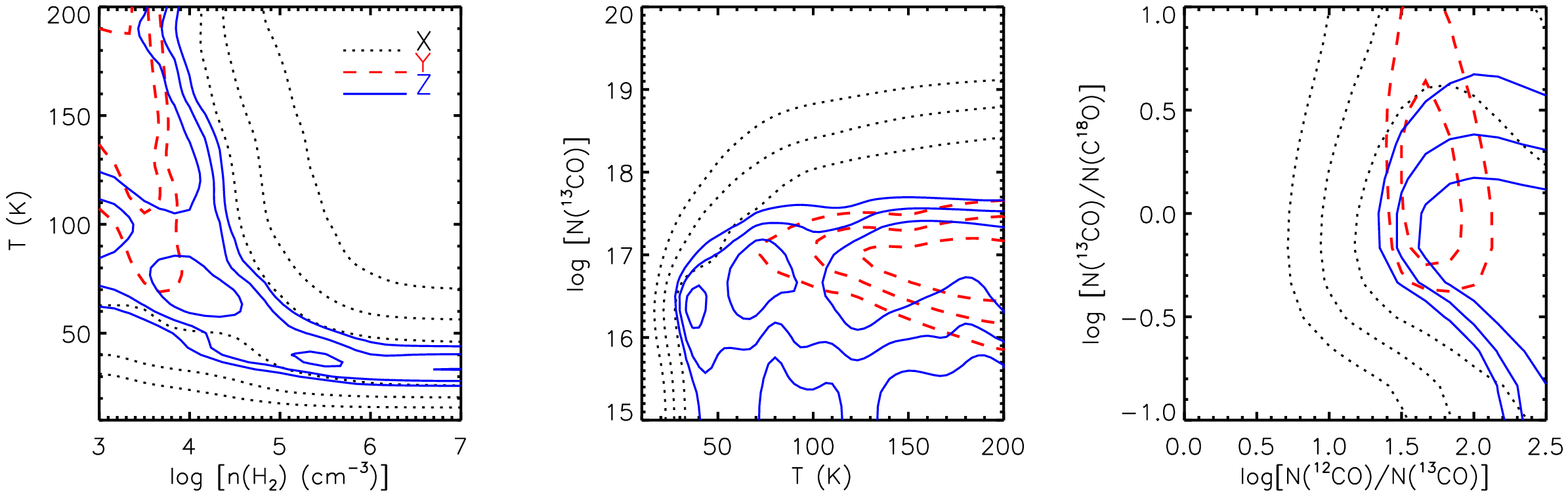,width=8in,angle=90}}
\caption{ Parameter space distributions of the best-fitting models in
  $X$, $Y$ and $Z$ components for different combinations of
  parameters. The contours represent the models lying within $\Delta
  \chi^{2}$\,=\,1, 2 and 3$\sigma$, which for six degrees of freedom
  corresponds to models within $\Delta \chi^{2}$ of 7.0, 12.8 and 20.0
  respectively. These fits to the kinematically decomposed spectral
  line fluxes provide better constraints on the parameter space than
  those using the integrated emission. The different components are
  represented with different colours and linestyles and it can be seen
  that the different components occupy different regions of parameter
  space, although there is overlap between all of them. The $X$
  component is clearly the least constrained as we have fewer line
  detections in this component. }
\label{fig:dec_param}
\end{figure*}

In Fig.~\ref{fig:dec_sleds} we show the best LVG model fits to the
individual kinematic components in SMM\,J2135. The corresponding
best-fit parameters and possible ranges of values within $\Delta \chi
^{2}=1\sigma$ of the best-fit are listed in Table~3. In
Fig.~\ref{fig:dec_param} we show the parameter space of our five model
parameters with likelihood contours that represent the models that lie
within $\Delta \chi^2$\,=\,1, 2 and 3$\sigma$ of the best-fit.  We can
better constrain the physical parameters when we fit to each kinematic
component individually. The different components occupy different
regions of parameter space in temperature, density and column
densities. We now discuss the properties of the individual kinematic
components:

\begin{table*}
\caption{LVG model parameters}
\begin{center} 
\smallskip
\begin{tabular}{llllllll}
\hline
\hline
\noalign{\smallskip}
Component & Method & $\log\frac{N(^{13} \rm CO)}{N(\rm C^{18}O)}$  &
$T_{\rm K}$ & $n$(H$_2$) & $N(^{12}$CO) & $\log\frac{N(^{12}\rm CO)}{N(^{13}\rm CO)}$ & $\chi^2$(reduced)  \\
          &        &              & (K)     & (cm$^{-3}$) & (cm$^{-2}$) &  \\ 
\hline\hline
Search ranges: & ... & 0.1--1   & 10--200 & $10^{3-7}$ & $10^{15-20}$ & 10$^{0-2.5}$ & ... \\ 
\hline
Cold & 2-phase     & 0.5 ($-$1--1)      & 50 (10--50)   &$10^3$ ($10^{3-7}$)   & $10^{19.5}$ ($10^{18-20}$)      & 2 (0.5--2.5)  & 4.1 \\
Warm & 2-phase    & 0 ($-$1--1)        & 90 (60--160) & $10^4$ ($10^{3.5-7}$)   &  $10^{20}$ ($10^{18.5-20}$)     & 2 (1.5--2.5)  & 4.1 \\
\hline\hline
$X$          & 3-component     & 0 ($-$0.5--0.5)        & 200 (30--200)    & $10^{3}$ (10$^{3-7}$)     &  $10^{19}$ (10$^{15-20}$)        & 2 (1.5--2.5)    & 1.8 \\
\hline
$Y$          & 3-component     & 0.5 (0--0.5)       & 140 (110--200)        & $10^{3.5}$ ($10^{3-3.5}$)   &  $10^{18.5}$ ($10^{18-18.5}$)  & 1.5 (1.5) & 2.5 \\ 
 \hline
$Z$          & 3 component     & 0 ($-$0.5-1)        & 140 (30-200)        & $10^{3.5}$ ($10^{3-7}$) &  $10^{19}$ ($10^{18-19.5}$)   & 2 (1.5--2.5)        & 3.4 \\

\hline\hline
\label{tab:LVGparam}
\end{tabular}
\end{center}
\footnotesize{ Note: For kinetic temperature ($T_{\rm K}$), H$_{2}$ density
  ($n$(H$_2$)) and column density ($N(^{12}$CO)) we give the parameter
  values associated with the minimum $\chi^2$ and the range in these
  parameters (within $\Delta\chi^2=1\sigma$) is given in brackets. The
  $\chi^2$ of the two-phase fit is based on 15 data
  points and 12 free parameters therefore to calculate
  the reduced $\chi^2$ we divide by 3 degrees of
  freedom. Similarly for the kinematically decomposed models the reduced $\chi^2$ is calculated
  by dividing by 9 degrees of freedom (15 data points -- 6
  free parameters). }
\end{table*}

%
%

%
%

{\it X-component:} This component has a velocity offset from the
dynamical centre of the system by +400\,km\,s$^{-1}$. Unfortunately,
as a result of it only being detected in $^{13}$CO(3--2) and
C$^{18}$O(5--4) the ISM properties we derive for this component are
very poorly constrained. Broadly, it has a temperature
which is only constrained to lie between 30--200\,K, a density range of
$n$(H$_2$)\,=\,10$^{3-7}$\,cm$^{-3}$ and the same high abundance ratio
of $N(^{13}$CO)/$N($C$^{18}$O)\,=\,1 as for the $Z$-component.

{\it Y-component:} This component is the centre of mass of the system,
lying at +165\,km\,s$^{-1}$. It is present in all the observed
emission lines in our dataset but it is marginally stronger in the
$^{13}$CO than in the C$^{18}$O emission lines. The LVG models suggest
that the $Y$-component exhibits the lowest abundance ratio of all
three components for $N(^{12}$CO)/$N(^{13}$CO)\,$\sim30$ and the
highest for $N(^{13}$CO)/$N($C$^{18}$O)\,$\sim1-3$, possibly
suggesting a relatively high abundance for $^{13}$CO. It is reasonably
well constrained to have a warmer best-fit solution than the $X$ and
$Z$ components ($T=110-200$\,K). The low abundance could be indicating
that component $Y$ is in a later stage of star formation, since
$^{13}$C are secondary nuclei produced in longer lived
low-to-intermediate mass stars. In such a scenario this component may
be the oldest of the three components. This is demonstrated by the
different abundance ratio vectors in Fig.~\ref{fig:lineratio}. We note
that the $Y$-component SLED fits tend to overpredict the $^{13}$CO and
underpredict the C$^{18}$O SLEDs and the best-fit does not peak at a
high enough excitation in C$^{18}$O.

{\it Z-component:} The $Z$-component is at the opposite edge of the
gas disk from the $X$-component at a velocity of
$\sim-70$\,km\,s$^{-1}$. It is very similar to the $X$-component in
terms of $T$, $n($H$_2)$ and abundance ratios. We observe it to
  be very weak in $^{13}$CO but strong in C$^{18}$O, in particular the
  highest rotational energy, C$^{18}$O(5--4). It also exhibits a high
abundance ratio of $N(^{12}$CO)/$N(^{13}$CO)\,$\sim100$ and a low
abundance ratio of $N(^{13}$CO)/$N($C$^{18}$O)\,$\sim1$. The low
$N(^{13}$CO)/$N($C$^{18}$O) ratio appears to be due to a genuinely
high abundance of C$^{18}$O. A similarly high abundance is found in
the $X$-component. We note that for the $Z$-component, the best-fit
model peaks at J$_{\rm up}\sim4$ whereas the data peaks closer to
J$_{\rm up}=5$ and the fit to the C$^{18}$O(5--4) is underpredicted by
$\sim3\sigma$, so the temperature is potentially even higher. The
properties of the $Z$-component are consistent with a warm, dense
medium undergoing preferentially massive star formation. The similar
$X$- and $Z$-components thus appear to be chemically younger than the
$Y$-component and they may be similar structures.

The $X$ and $Z$-components require a lower abundance ratio of
[$^{13}$CO]\,/\,[C$^{18}$O] than the $Y$-component and this abundance
ratio is surprisingly low ($<1$), implying enhanced C$^{18}$O
abundance in the outer regions of the galaxy. Whereas the
$Y$-component exhibits the lowest ratio of $N(^{12}$CO)/$N(^{13}$CO)
out of the three components, potentially implying an enhanced
$^{13}$CO abundance in the dynamical centre of the galaxy.

\section{Discussion}
\label{sec:discuss}
Our measurements of optically thin $^{13}$CO and C$^{18}$O in the ISM
of SMM\,J2135 suggest a molecular gas mass of $M_{\rm
  gas}\sim(0.8-1.5)\times10^{10}$\,M$_{\odot}$ (see
\S~\ref{sec:gasmass}).  When combined with
the stellar mass, $M_{*}=(3.0\pm1)\times10^{10}$M$_{\odot}$, the
implied dynamical mass ($M_{\rm dyn}$\,=\,$M_{\rm gas}+M_{\star}$) is
within 2\,$\sigma$ of that derived from the galaxy dynamics $M_{\rm
  dyn}=(6.0\pm0.5)\times10^{10}$M$_{\odot}$ \citep{Swinbank11}, if we
take the upper limit on our derived gas mass.  We therefore derive a
ratio of $M_{\rm gas}$\,/\,$M_{\rm dyn}\sim0.25$ which is comparable
to the average molecular gas fractions found for high-redshift
star-forming galaxies of $M_{\rm gas}$\,/\,$M_{\rm dyn}\sim
$\,0.3--0.5 at $z\sim $\,1--2 (\citealt{Tacconi13};
\citealt{Bothwell13}).  Using $M_{\rm
  gas}\sim1.5\times$\,10$^{10}$\,M$_{\odot}$ we derive $\alpha_{\rm
  CO}\sim $\,0.9 which is close to the canonical value for ULIRGs
(i.e. \citealt{Downes98}) but lower than our previous estimate of
$\alpha_{\rm CO}$\,=\,2 in \cite{Danielson11}.

Modelling the $^{12}$CO, $^{13}$CO and C$^{18}$O SLEDs,  we find that
the ISM is best described by a two-phase ISM comprising a `cold'-phase
($T_{\rm K}\sim50$\,K) tracing the extended gas disk, and a
`hot'-phase ($T_{\rm K}\sim90$\,K), with an average density of
$\sim10^{3-4}$\,cm$^{-3}$ (presumably tracing the denser star
forming regions).  However, due to the strong differences in $J_{\rm
  peak}$ for the $^{13}$CO and C$^{18}$O SLEDs we find that a
two-phase fit to the integrated emission is unable to adequately
describe both simultaneously.

However, SMM\,J2135 exhibits highly-structured molecular emission
spectra, and previous work has demonstrated that the source can be
kinematically decomposed into three main components which are
coincident with the bright star-forming clumps visible in the
high-resolution observed 870$\mu$m SMA and JVLA $^{12}$CO(1--0) maps
(\citealt{Swinbank11}). Thus, deriving the properties of such a
complex system via the integrated flux, although illustrative for
comparison to both local and high-redshift observational studies, is
not probing the true conditions in the star-forming regions
themselves. Furthermore, using optically thick gas tracers gives a
biased view of the properties of the cold molecular gas and does not
probe the densest regions in which the stars are forming. Thus, we fit
a three-component kinematic model ($X$, $Y$, $Z$) to the spectra of
the two lower abundance (and thus typically optically thinner) gas
tracers: $^{13}$CO and C$^{18}$O, in order to provide a more detailed
understanding of the ISM conditions within this high-redshift
star-forming galaxy. By kinematically decomposing the SLEDs into their
constituent components ($X$, $Y$ and $Z$) we show that there are no
strong differences in the excitation of the $^{13}$CO and C$^{18}$O in
any of the components, although we do find abundance gradients between
the components.

\cite{Danielson11} proposed that the highly structured $^{12}$CO
emission line morphology of SMM\,J2135 arises from a merging system in
which components $Y$ and $Z$ correspond to two different interacting
galaxies with the $X$-component being a diffuse tidal feature.
However, high-resolution ($\sim $\,0.2$"$) dynamical maps of the
$^{12}$CO(6--5) and $^{12}$CO(1--0) emission showed that the bulk of
the gas in fact lies in a 5\,kpc diameter, rotationally supported,
clumpy disk \citep{Swinbank11}.  In this model the highly-structured
$^{12}$CO emission arises from a number of bright clumps within the
disk.

Our new $^{13}$CO and crucially, C$^{18}$O observations are consistent
with the latter interpretation: the enhanced C$^{18}$O is found in the
outer regions of the disk (the high C$^{18}$O is typically associated
with young, high mass star-formation), and the enhanced $^{13}$CO
(generally associated with older, lower mass stars), is located in the
central regions.  However, we note that the gas disk has a very low
Toomre $Q$, ($Q$\,=\,0.50\,$\pm$\,0.15; \citealt{Swinbank11}),
suggesting that a significant, major accretion event must have
recently occurred and the star formation has yet to stabilise the disk
back to $Q$\,=\,1 (e.g.~\citealt{Hopkins12}).  One possible local
analogue of SMM\,J2135 is the luminous galaxy NGC\,6240
(e.g. \citealt{Tacconi99}; \citealt{Iono07}).  This system has been
found to comprise a thick, highly turbulent disk, centred between the
two nuclei of the merging progenitors.  In this system, the tidally
stripped gas from the two progenitors has settled into a rotationally
supported disk in the potential well of the remnant.  The variations
in abundance across SMM\,J2135 may imply that it is in a similar
regime, since such chemical variations and low Toomre $Q$ are unlikely
to arise through the gravitational collapse of a single disk.

\subsection{Cosmic rays as a heating source}
\label{sec:cosmic}
There is evidence that in some star-forming galaxies, heating from
cosmic rays may play a more important role than photons, due to their
ability to penetrate and volumetrically heat dense gas
(e.g. \citealt{Goldsmith78}; \citealt{HD08}; \citealt{Bradford03};
\citealt{Bayet11a}; \citealt{Bayet11b}; \citealt{PPP12c}). Given the
high star formation rate of this galaxy
(SFR\,$\sim400$\,M$_{\odot}$\,yr$^{-1}$), particularly in the clumps, we
now look at the energetics of the heating and cooling of the cold
dense gas in this system, comparing the potential contributions from
cosmic ray and UV heating to the total cooling we expect from
our CO lines (one of the main coolants in the cold ISM).

\subsubsection{Cosmic Ray Heating Rate}
Following \cite{Suchkov93} and \cite{Bradford03}, assuming that cosmic
rays are produced at a rate proportional to the supernova rate, and
hence the current star formation rate, and are removed from
star-forming regions by high-velocity galactic winds, we derive the
cosmic ray heating rate, $\chi_{H_2}$, per H$_2$ molecule:
\begin{equation}
\chi_{\rm H_2}=\zeta_{\rm p}\Delta Q (\rm erg s^{-1} per \, H_2),
\end{equation}
where, $\zeta_{\rm p}$ is the ionisation rate per H$_2$ molecule and $\Delta
Q$ is the thermal energy deposited per ionisation (17--20\,eV;
\citealt{Goldsmith78}). By scaling from the Milky Way we determine the
ionisation rate to be:
\begin{equation}
\zeta_{\rm p, SMMJ2135}=\zeta_{\rm p, MW}\times\frac{\psi_{\rm SMMJ2135}}{\psi_{\rm MW}}\times\frac{v_{\rm MW}}{v_{\rm w, SMMJ2135}},
\end{equation}
where $\zeta_{\rm p, MW}$ is the local Galactic ionisation rate
(2--7$\times10^{-17}$s$^{-1}$ for dense gas;
e.g. \citealt{Goldsmith78}; \citealt{vandishoeck86}); $\psi_{\rm
  SMMJ2135}$ is the supernova rate per unit area in SMM\,J2135;
$\psi_{\rm MW}$ is the supernova rate per unit area (over the entire
disk) of the Milky Way; $v_{\rm w}$ is the cosmic ray diffusion
velocity from the Galactic disk and $v_{\rm w, SMMJ2135}$ is the wind
velocity of SMM\,J2135. We assume a Galactic star formation rate of
0.68--1.45\,M$_{\odot}$\,yr$^{-1}$ across the whole disk (the upper
limit taken from \citealt{Robit10}), resulting in a SFR surface
density of $\sum_{\rm
  SFR,MW}=(9.6-20.1)\times10^{-4}$\,M$_{\odot}$\,yr$^{-1}$\,kpc$^{-2}$
and we assume the cosmic ray diffusion velocity from the Galactic disk
to be $v_{\rm w}=10$\,km\,s$^{-1}$ \citep{Suchkov93}.

Previous observations have suggested that $>$\,50\% of the star
formation in SMM\,J2135 may be occurring in the clumps, which appear
to be closely associated with the kinematic components $X$, $Y$ and
$Z$ (\citealt{Swinbank11}). To search for cosmic ray heating we
therefore concentrate on these regions. We assume the supernova rate
per unit area $\psi_{\rm SMMJ2135}$ to be proportional to the SFR
surface density in the clumps of
SFR\,=\,30\,--\,90\,M$_{\odot}$\,yr$^{-1}$ (assuming a Salpeter
initial mass function; \citealt{Swinbank11}). We also assume clump
radii of 100\,--\,200\,pc, and that the stars do not migrate far from
the clumps before they become supernovae. We adopt a wind velocity in
SMM\,J2135 of $v_{\rm w, SMMJ2135}$\,=\,200\,km\,s$^{-1}$,
approximately the FWHM of the individual components. Typical values
for wind velocity in local starbursts and high-redshift submillimetre
galaxies are $v_{\rm w}=(1-3)\times10^3$\,km\,s$^{-1}$
(\citealt{Banerji11}) so $v_{\rm w, SMMJ2135}=200$\,km\,s$^{-1}$ is
likely to be a lower limit (thus an upper limit on the cosmic ray
heating rate).  This suggests an ionisation rate of $\zeta_{\rm p,
  SMMJ2135}\sim(1-100)\times10^{-13}$s$^{-1}$.  Assuming $\Delta
Q=20$\,eV per ionisation, the power deposited per H$_2$ molecule is
$\chi_{\rm H_2}\sim(30-3300)\times10^{-25}$\,erg\,s$^{-1}$. This is
higher than the cosmic ray heating rate of $\chi_{\rm
  H_2}=(5-18)\times10^{-25}$\,erg\,s$^{-1}$ in the local starburst
galaxy NGC\,253 \citep{Bradford03}. However, given that the average
star formation rate is significantly higher in SMM\,J2135 than in
NGC\,253 (SFR$\sim2-3$\,M$_{\odot}$\,yr$^{-1}$; \citealt{Ott05}), it
would be expected that the ionisation rate from cosmic rays should be
significantly higher.

\cite{PPP10c} predicts high temperatures of $>$80--240\,K in compact
starbursts with cosmic ray ionisation rates of
$(5-20)\times10^{-14}$s$^{-1}$. In SMM\,J2135 we derive a
significantly higher ionisation rate (up to $50\times$ higher) and
furthermore, the best-fit temperatures we derive using LVG modelling
for the individual components $X$, $Y$ and $Z$ range between
$140-200$\,K. Thus these high temperatures could also hint that the
clumps may be dominated by cosmic ray heating.

\subsubsection{UV Heating Rate}
For comparison to the cosmic ray heating we also estimate the likely
UV photon heating per H$_{2}$ molecule from massive star formation
assuming that the far-infrared luminosity is dominated by OB
stars. The typical efficiency of photoelectric heating in PDRs is
$\sim$\,0.3\% (i.e. \citealt{Weingartner06}), therefore the UV heating
input per H$_{2}$ molecule can be estimated using,
$\chi_{H_2}=0.003L_{\rm IR}$/($M_{\rm H_2}/m_{\rm H_2}$). Given an
intrinsic infrared luminosity (8--1000$\mu$m) of $L_{\rm
  IR}=2.3\times10^{12}$L$_{\odot}$ (\citealt{Ivison10eyelash}), we
derive a heating rate per H$_{2}$ molecule of $\chi_{\rm
  H_2}\sim15\times10^{-25}$erg\,s$^{-1}$ over the whole system. Again,
it is likely that the heating rate in the clumps is higher than the
heating rate integrated over the whole system.

To estimate the UV heating per clump we assume an average clump mass
of $\sim$1.3$\times10^9$M$_{\odot}$ (\citealt{Danielson11}) and that
each clump contributes $\sim$1/8th of the observed $L_{\rm IR}$
($\sim3\times10^{11}$L$_{\odot}$), which is the typical clump fraction
from the total 260\,$\mu$m restframe emission
(\citealt{Swinbank10Nature}). The heating rate per H$_2$ molecule can
then be estimated to be $\chi_{\rm
  H_2}\sim45\times10^{-25}$erg\,s$^{-1}$. This is similar to the
minimum heating per clump that could be provided by cosmic rays, hence
we conclude that both UV and cosmic rays are energetically capable of
heating the gas.

However, we can take this estimate one stage further and ask what is
the necessary distribution of gas needed for PDRs to heat the whole
reservoir of warm gas?  If the warm gas in the system is being heated
by PDRs, then all that gas has be to in the vicinity of a hot star to
reach a far-UV radiation field of $G_0\sim1\times10^{3-3.6}$ Habing
fields (\citealt{Ivison10eyelash}; \citealt{Danielson11}). Again,
using the far-infrared luminosity of $2.3\times10^{12}$L$_{\odot}$ and
making the approximation that the UV radiation field is dominated by
O5 stars (luminosity $\sim8\times10^5$\,L$_{\odot}$ per star), results
in $\sim3\times10^{6}$ O5 stars in SMM\,J2135. Using the SED of an O5
star, in order to have $G_0 \ge 1\times10^3$, the gas has to be within
3\,pc of the star. So, using spheres of radius $\sim3$\,pc, assuming
$\sim3\times10^{6}$ O5 stars, and using the derived gas density, we
can calculate the gas mass exposed to $G_0 \ge 1\times10^3$. For our
minimum density of $n$(H$_2)\sim\times10^3$\,cm$^{-3}$ (\S\ref{sec:int_sleds}),
this becomes $M\sim1.6\times10^{10}$\,M$_{\odot}$ which is in reasonable
agreement with our estimated gas mass from $^{13}$CO
(\S\ref{sec:gasmass}). However, the gas can only be heated by UV
up to $A_V\sim5$ mag from the star (e.g., \citealt{Tielens85},
Fig. 7). For densities up to $2\times10^{3}$\,cm$^{-3}$, the whole
$\sim3$\,pc radius sphere is in the $A_V < 5$ mag region. Thus, for
gas densities between $(1-2)\times10^{3}$\,cm$^{-3}$, the amount of
PDR-heated gas will be between
$M=(1.5-3.0)\times10^{10}$\,M$_{\odot}$. If much of the gas is at
significantly higher density then PDRs will not be able to provide
enough heating, since only the surface layer is heated. For example,
for $n$(H$_2$)$ \sim1\times10^5$\,cm$^{-3}$, $A_V = 5$ is reached at
only $\sim$\,0.3\,pc from the star and the amount of gas that can be heated
is only $M\sim1.6\times10^8$\,M$_{\odot}$.  Therefore, we can conclude that
if the bulk gas density is significantly higher than
$n$(H$_2$)$\sim2\times10^3$\,cm$^{-3}$ (which indeed we may be observing
in some of the star-forming regions), then PDRs will not be able to
provide the heating, and we need a mechanism that is capable of
volumetrically heating the gas, such as cosmic rays.

\subsubsection{Cooling Rate}
To estimate the cooling rate per H$_2$ molecule, we calculate the
total CO luminosity in all species following
\cite{SolomonVandenBout05} and \cite{HD08}:
\begin{equation}
\Sigma L_{\rm CO}=\Sigma 1.04\times10^{-3}S_{\rm CO,i}\Delta V \nu_{\rm rest,i}D_L^2(1+z)^{-1}/\mu,
\end{equation}
where $\mu$ is the magnification factor of $37.5\pm4.5$ and D$_L$ is
the luminosity distance in Mpc. We estimate a total intrinsic CO
luminosity of $L_{\rm CO}\approx1.8\times10^{8}$\,L$_{\odot}$. Using
our cold gas mass estimated from $^{13}$CO(3--2) of $M_{\rm
  gas}=1.5\times10^{10}$\,M$_{\odot}$, our cooling rate per H$_2$
molecule is $L_{\rm CO}/(M_{\rm H_2}/$\,m$_{\rm
  H_2})\sim0.8\times10^{-25}$\,erg\,s$^{-1}$. In this calculation we
have not included the [C{\sc ii}]\,157.8\,$\mu$m emission which is the
dominant coolant in the outer envelopes of molecular clouds. Including
the [C{\sc ii}]\,157.8\,$\mu$m in our calculation our cooling per
H$_{2}$ becomes $\chi_{\rm H_2}\sim20\times10^{-25}$\,erg\,s$^{-1}$.
However, at high A$_{\rm V}$ (A$_{\rm V}>5$) the [C{\sc
  ii}]\,157.8\,$\mu$m abundance significantly decreases and so has a
minimal contribution to the cooling in these regions.

\subsubsection{Balancing the Temperature}
We have derived heating rates from cosmic rays and UV photons of
(30--3300)$\times10^{-25}$\,erg\,s$^{-1}$ and
$\sim45\times10^{-25}$\,erg\,s$^{-1}$ respectively and we derive
cooling rates from the atomic and molecular line emission of
  $\sim(0.8-20)\times10^{-25}$\,erg\,s$^{-1}$.  Although crude,
overall this shows that both the cosmic ray and UV heating rates are
comparable to the cooling rate. However, we also demonstrate that for
densities of $n$(H$_2$)$>2 \times10^5$\,cm$^{-3}$, UV heating (from O5
stars) would not provide enough heating, due to extinction.

Moreover, \cite{PPP12c} have suggested that in
regions of enhanced cosmic ray density, the cosmic rays are able to
penetrate to the core of the dense gas and volumetrically raise the
temperature of the star-forming cores, which sets new initial
conditions for star-formation and increases the characteristic mass
of young stars.

Therefore, although we estimate that both the UV and cosmic rays are
capable of balancing the cooling rate in SMM\,J2135, the high gas
densities and the high ($>100$\,K) temperatures we derive for the
star-forming clumps may provide indirect evidence that cosmic ray
heating is particularly important in the individual kinematic
components of SMM\,J2135. It is therefore possible that we are seeing
evidence of this where we see an enhancement of C$^{18}$O, which is
also indicative of preferentially massive star formation, possibly due
to a raised initial temperature for star formation in these regions.

Finally, it is important to note that as well as cosmic ray heating
and photon heating discussed here, X-ray heating, shocks and turbulent
heating have been found to play an important role in the heating of
interstellar gas both in the Milky Way and in other
systems. \cite{Swinbank11} finds a highly turbulent ISM in SMM\,J2135
and turbulence can have a similar effect to cosmic rays of
volumetrically heating the gas (i.e. \citealt{Meijerink11};
\citealt{PPP12c}; \citealt{Ao13}; \citealt{Meijerink13}).

\section{Conclusions}
\label{sec:conc}

We analyse observations of $^{13}$CO and C$^{18}$O emission from the
lensed, $z$\,=\,2.3 ULIRG, SMM\,J2135.  We have
combined these observations with our previous $^{12}$CO measurements
to better constrain the conditions in the ISM of this system by
analysing the galaxy-integrated fluxes and the kinematically
decomposed emission. Using these lower abundance tracers of H$_2$ we
have been able to remove some of the degeneracies in modelling the ISM. We
summarise our conclusions as:

\begin{enumerate}
\item{We demonstrate that the $^{13}$CO emission is likely to be
    optically thin ($\tau\ll 1$) and we use this to estimate the total
    cold gas mass of $M_{\rm gas}\sim1.5\times10^{10}$M$_{\odot}$
    which is consistent with the dynamical and stellar limits on the
    total mass of the system. This implies $\alpha_{\rm
      CO}\sim0.9$ for this high-redshift ULIRG.}
\item{We detect C$^{18}$O and measure a surprisingly high flux
    resulting in a $^{13}$CO/C$^{18}$O flux ratio $\sim$4--15$\times$
    lower than that measured in the Milky Way. Since $^{18}$O is
    associated with the winds from massive stars, it is possible that
    this enrichment of C$^{18}$O may be due to the presence of
    preferentially massive star formation.}
\item{The ISM is best described by a two-phase model; a `cold' phase
    at $\sim$50\,K with a density of $n$(H$_2$)\,$\sim$\,10$^3$cm$^{-3}$, and a
    `hot' phase at $\sim$90\,K and $n$(H$_2$)\,$\sim$\,10$^4$cm$^{-3}$
    respectively. However, the SLEDs of $^{13}$CO and C$^{18}$O peak
    at different J$_{\rm up}$ despite both appearing to be optically
    thin. We attribute this to variations of [$^{13}$CO]/[C$^{18}$O]
    within the galaxy and therefore kinematically decompose the line
    emission from the system into three main components, $X$, $Y$ and
    $Z$. We find that the $Y$-component, which appears to be
    coincident with the centre of mass of the system, is warm ($T_{\rm
      K}\sim 140$\,K) and dense ($n$(H$_2$)\,$\sim$\,10$^{3.5}$cm$^{-3}$)
    potentially with a higher abundance of $^{13}$CO
    ([$^{12}$CO]/[$^{13}$CO]$\sim30$) implying older star formation
    from intermediate mass stars. In contrast, the $Z$- and
    $X$-components are similar to each other. Both these regions
    display low [$^{13}$CO]/[C$^{18}$O] ratios possibly implying
    enhanced massive star formation leading to a higher abundance of
    C$^{18}$O in these regions.}

\item{We have derived an average cooling rate from all the observed CO
    lines of $\sim(0.8-20)\times10^{-25}$erg\,s$^{-1}$ per H$_2$
    molecule. We determine the possible contribution to the heating
    from cosmic rays (originating largely in supernovae) and from UV
    photon heating of (30--3300)$\times10^{-25}$erg\,s$^{-1}$ and
    $\sim45\times10^{-25}$erg\,s$^{-1}$ respectively. Although crude,
    both cosmic ray heating and UV heating can plausibly balance the
    cooling occurring in the system. However, the high temperatures
    ($T_{\rm K}=140-200$\,K) derived in the highest density components
    may suggest that cosmic rays may play a more important role than
    UV heating in this system.}
\end{enumerate}

Since SMM\,J2135 is a representative high-redshift ULIRG, this study
paves the way for future detailed studies of this population. With
ALMA in full science operations we will be able to resolve sub-kpc
structure in these faint lines and derive resolved ISM
properties. Furthermore, ALMA will enable the important observations
of high density chemical tracers such as CS and HCN, giving insight
into the origin of the intense star formation activity in this galaxy
and potentially testing for systematic variation in the relative
abundances of $^{12}$CO, $^{13}$CO and C$^{18}$O which may reflect
variation in the IMF.

\section*{acknowledgments}

We would like to thank the anonymous referee for a thorough and
constructive report which significantly improved the content and
clarity of this paper. ALRD acknowledges an STFC studentship
(ST/F007299/1). AMS gratefully acknowledges an STFC Advanced
Fellowship through grant ST/H005234/1. IRS acknowledges support from
STFC, a Leverhulme Fellowship, the ERC Advanced Investigator programme
DUSTYGAL 321334 and a Royal Society/Wolfson Merit Award.  We thank
Francoise Combes and Steve Hailey-Dunsheath for useful conversations
and Padelis Papadopoulos for extensive comments and useful
suggestions.  We thank John Helly for his help. We thank the IRAM
staff Melanie Krips and Roberto Neri for help provided during the
observations and for data reduction guidance. The research
  leading to these results has received funding from the European
  Commission Seventh Framework Programme (FP/2007-2013) under grant
  agreement No 283393 (RadioNet3). The raw data from JVLA and IRAM on
  which this analysis is based can be accessed through the JVLA
  archive (programme code 11B-062) and through contacting IRAM with
  the programme code U0B6.


\end{document}

%% file: line_ratios_combined.tex
\begin{tabular}{c c c c c c }
\hline\hline
Species or &  $\nu_{\rm rest}$ & Integrated & \multicolumn{3}{c}{Kinematically decomposed flux$^{a,b,c}$}  \\
line ratio &                  & flux &   $Z$ & $Y$ & $X$  \\
           &     (GHz)        & (Jy\,km\,s$^{-1}$) & \multicolumn{3}{c}{(Jy km s$^{-1}$)}\\
\hline\hline
$^{13}$CO(1--0)  & 110.2014 & $<0.07$       & $<0.03$ & $<0.05$ & $<0.03$   \\
$^{13}$CO(3--2)  & 330.5880 & $0.66\pm0.08$ & $0.11\pm0.06$ & $0.43\pm0.09$ & $0.09\pm0.06$\\
$^{13}$CO(5--4)  & 550.9263 & $0.38\pm0.14^{e}$ & $<0.43$       & $0.46\pm0.07$ & $<0.21$ \\
$^{13}$CO(7--6)  & 771.1841 & $<0.41$       & $<0.26$       & $<0.38$       & $<0.26$ \\
\hline
C$^{18}$O(1--0)   & 109.7822 & $<0.07$ &$<0.03$ &$<0.05$  & $<0.03$ \\
C$^{18}$O(3--2)   & 329.3305 & $0.41\pm0.08$  & $0.16\pm0.06$ & $0.28\pm0.09$ & $<0.16$ \\
C$^{18}$O(5--4)   & 548.8310 & $0.87\pm0.16$  & $0.43\pm0.08$ & $0.34\pm0.13$ & $0.21\pm0.09$ \\
C$^{18}$O(7--6)   & 768.2514 & $<0.41$        & $<0.26$       & $<0.38$       & $<0.26$ \\
\hline
$^{12}$CO(1--0)$^{d}$  & 115.2712 & $2.16\pm 0.11$ & $0.56\pm0.11$ & $1.3\pm0.2$ & $0.4\pm0.2$  \\
$^{12}$CO(3--2)$^{d}$  & 345.7959 & $13.20\pm0.10$ & $3.6\pm0.6$ & $7.6\pm1.1$ & $2.3\pm1.1$  \\
$^{12}$CO(4--3)$^{d}$  & 461.0408 & $17.3\pm1.2$   & $3.8\pm0.8$ & $9.9\pm1.4$ & $4.0\pm1.4$  \\
$^{12}$CO(5--4)$^{d}$  & 576.2679 & $18.7\pm0.8$   & $7.2\pm0.7$ & $7.1\pm1.3$ & $4.5\pm1.3$  \\
$^{12}$CO(6--5)$^{d}$  & 691.4731 & $21.5\pm1.1$   & $8.2\pm0.9$ & $8.7\pm1.6$ & $5.1\pm1.6$  \\
$^{12}$CO(7--6)$^{d}$  & 806.6518 & $12.6\pm0.6$   & $4.7\pm0.5$ & $5.9\pm0.9$ & $1.1\pm0.9$  \\
$^{12}$CO(8--7)$^{d}$  & 921.7997 & $8.8\pm0.5$    & $3.2\pm0.3$ & $3.1\pm0.6$ & $2.5\pm0.6$  \\

\hline
$^{12}$CO(1--0)/C$^{18}$O(1--0) & ... & $>31$ &  $>19$ & $>26$ & $>13$\\
$^{12}$CO(3--2)/C$^{18}$O(3--2) & ... & $32\pm6$ & $23\pm9$ & $27\pm10$  & $>15$ \\
$^{12}$CO(5--4)/C$^{18}$O(5--4) & ... & $21\pm4$ & $16\pm4$ & $21\pm15$ & $21\pm17$ \\
$^{12}$CO(7--6)/C$^{18}$O(7--6) & ... & $>31$        & $>18$  & $>16$ & $>4$ \\
\hline      
$^{12}$CO(1--0)/$^{13}$CO(1--0) & ... & $>31$ &  $>19$ & $>26$ & $>13$\\                   
$^{12}$CO(3--2)/$^{13}$CO(3--2) & ... & $20\pm2$ & $32\pm17$ & $18\pm5$ & $25\pm21$ \\
$^{12}$CO(5--4)/$^{13}$CO(5--4) & ... & $49\pm18$ & $>17$ & $15\pm10$ & $>21$ \\
$^{12}$CO(7--6)/$^{13}$CO(7--6) & ... & $>31$        & $>18$  & $>16$ & $>4$ \\
\hline                            
$^{13}$CO(3--2)/C$^{18}$O(3--2) & ... & $1.6\pm0.4$  & $0.69\pm0.46$ & $1.5\pm0.6$ & $>0.56$ \\
$^{13}$CO(5--4)/C$^{18}$O(5--4) & ... & $0.44\pm0.18$ & $<1.0$ & $1.3\pm0.5$ & $<1.0$ \\
\hline
\hline
\end{tabular}